\documentclass[twocolumn]{aastex62}

\usepackage{apjfonts}

\shorttitle{Ionized Nebulae Around Two New Symbiotic Stars}
\shortauthors{Bond et al.}

\newcommand{\Ha}{H$\alpha$}
\newcommand{\Hb}{H$\beta$}
\newcommand{\Hg}{H$\gamma$}

\newcommand{\kms}{{\>\rm km\>s^{-1}}}
\newcommand{\masyr}{\rm mas\, yr^{-1}}

\newcommand{\sprocess}{{\it s}-process}

\def\cai{\ion{Ca}{1}}
\def\caii{\ion{Ca}{2}}
\def\feii{\ion{Fe}{2}}
\def\hei{\ion{He}{1}}
\def\heii{\ion{He}{2}}

\def\nii{\ion{N}{2}}

\def\nai{\ion{Na}{1}}
\def\neiii{\ion{Ne}{3}}

\def\oii{\ion{O}{2}}
\def\oiii{\ion{O}{3}}

\def\ovi{\ion{O}{6}}
\def\sii{\ion{S}{2}}
\def\srii{\ion{Sr}{2}}

\def\Gaia{{\it Gaia}}
\newcommand{\GALEX}{{\it GALEX}}

\newcommand{\HST}{{\it HST}}

\newcommand{\IRAS}{{\it IRAS}}

\newcommand{\TESS}{{\it TESS}}

\def\D141{[D75]\,141} 

\def\GR{GR~Cyg}
\def\StDr{StDr~Objet~14}

\begin{document}

%\title{Ionized Nebulae Around the Late-Type Semiregular Variable Stars GR\,Cygni and [D75]\,141\footnote{Based in part on observations obtained with the Hobby-Eberly Telescope (HET), which is a joint project of the University of Texas at Austin, the Pennsylvania State University, Ludwig-Maximillians-Universit\"at M\"unchen, and Georg-August Universit\"at G\"ottingen. The HET is named in honor of its principal benefactors, William P. Hobby and Robert E. Eberly.} }

\title{Ionized Nebulae Around Two New Symbiotic Stars: GR\,Cygni and [D75]\,141\footnote{Based in part on observations obtained with the Hobby-Eberly Telescope (HET), which is a joint project of the University of Texas at Austin, the Pennsylvania State University, Ludwig-Maximillians-Universit\"at M\"unchen, and Georg-August Universit\"at G\"ottingen. The HET is named in honor of its principal benefactors, William P. Hobby and Robert E. Eberly.} }

\author[0000-0003-1377-7145]{Howard E. Bond}
\affil{Department of Astronomy \& Astrophysics, Pennsylvania State University, University Park, PA 16802, USA}
\affil{Space Telescope Science Institute, 
3700 San Martin Dr.,
Baltimore, MD 21218, USA}

%\author[0000-0003-2071-2956]{Soumyadeep Bhattacharjee}
%\affiliation{Department of Astronomy, California Institute of Technology, 1216 E. California Blvd, Pasadena, CA, 91125, USA}

\author[0009-0008-5193-4053]{Calvin Carter}
\affil{Rocket Girls Ranch Observatory,
7215 Paldao Dr.,
Dallas, TX 75240, USA }

\author[0009-0009-0300-3892]{Sven E. Eklund}
\affil{Eklund Smurf Observatory,
Villav\"agen 11,
S-791 37 Falun,
Sweden
}

\author[0009-0005-3715-4374]{Peter Goodhew}
\affil{Deep Space Imaging Network, 108 Sutton Court Rd., London, W4 3EQ, UK}

\author[0000-0001-6805-9664]{Ulisse Munari}
\affil{
INAF National Institute of Astrophysics, 
Astronomical Observatory of Padova, 
36012 Asiago (VI), Italy}

%\author[0000-0002-9018-9606]{Dana Patchick}
%\affil{Deep Sky Hunters Consortium, 1942 Butler Ave. Los Angeles, CA 90025, USA }

%\author[0009-0005-9964-1602]{Daniel Stern}
%\affil{MEA Observatory,
%16252 Andalucia Ln., 
%Delray Beach, FL  33446, USA }

\author[0009-0009-3986-4336]{Jonathan Talbot}
\affil{Stark Bayou Observatory, 1013 Conely Cir., Ocean Springs, MS 39564, USA}

%\author[0000-0002-4964-4144]{John R. Thorstensen}
%\affil{Department of Physics \& Astronomy, 6127 Wilder Laboratory, Dartmouth College, Hanover, NH 03755, USA}

\author[0000-0003-2307-0629]{Gregory R. Zeimann}
\affil{Hobby-Eberly Telescope, University of Texas at Austin, Austin, TX 78712, USA}

\correspondingauthor{Howard E. Bond}
\email{heb11@psu.edu}

\begin{abstract}

In the course of selecting targets for a spectroscopic survey of faint planetary-nebula nuclei, we encountered two anomalous cases of late-type variable stars surrounded by nebulosities discovered by amateur astronomers. We obtained follow-up deep narrow-band images of the two nebulae: \StDr\ (diameter $\sim$$1\farcm5$) and StDr~4 (diameter $\sim$$2\farcm$5). \StDr\ is found to be a low-excitation nebula, bright only in \Ha+[\nii], and showing a unique double-ring ``bull's-eye'' morphology. StDr~4 exhibits a bow-shock structure, bright in \Ha\ and [\sii], embedded in an extended emission nebula. Both central stars, respectively \GR\ and \D141, are found to be pulsating semiregular variable stars. Our spectroscopic observations show \GR\ to be a cool carbon star of type C9,2, with emission lines of the Balmer series, [\oii] and [\oiii], [\neiii], [\sii], and \feii. A synthetic narrow-band image of \GR\ shows that the source of its strong [\oii] $\lambda$3727 emission line is a knot displaced from the star by about $1\farcs6$, indicative of a recent collimated ejection event. \D141\ is a M6\,III red giant, with weak Balmer-series emission.  Both stars show a blue excess, and \GR\ was detected in the near-UV by the \GALEX\ spacecraft. These data establish both stars as new symbiotic binaries, containing a late-type red giant and a hot companion accreting from its stellar wind. We speculate that the double rings surrounding \GR\ were ejected during helium-shell thermal pulses (or conceivably from recurring nova outbursts), while the nebula around \D141\ may be the result of a collision of the symbiotic star's ejecta with an interstellar cloud. Our accidental discoveries of these two binaries because they are surrounded by faint nebulae supports the view that there is a significant population of ``missing'' symbiotic stars yet to be discovered.

\null\vskip 0.25in

\end{abstract}


% \keywords{blah --- blah}

\section{Introduction \label{sec:intro} }

In an ongoing survey, several of us and colleagues are obtaining spectroscopy of central stars of faint Galactic planetary nebulae (PNe). 
The project is carried out with the second-generation Low-Resolution Spectrograph (LRS2) of the 10-m Hobby-Eberly Telescope (HET; \citealt{Ramsey1998,Hill2021}), located at McDonald Observatory in west Texas, USA\null. 
Detailed information about the survey, and references to previous papers, can be found in our most recent publication \citep[Paper~VIII;][]{BondPaperVIII}. 
 
Our target selection emphasizes central stars of the numerous new PNe with very low surface brightnesses that have been 
discovered in recent years by amateur astronomers, mostly through systematic examination of publicly available sky surveys. Many of these new discoveries are followed up by advanced amateurs, using small telescopes with modern low-noise detectors to accumulate long exposure times. The resulting deep images often reveal remarkable nebular details. 

The {\tt planetarynebulae.net} website\footnote{\url{https://planetarynebulae.net}} (hereafter PNe.net), maintained by Pascal Le D\^u and Thomas Petit, is a leading resource used by amateurs to present their newly discovered PNe and followup imagery. Objects that appear to be true PNe, or strong candidates, are cataloged in the widely used online Hong-Kong/AAO/Strasbourg/H$\alpha$ Planetary Nebulae (HASH) database\footnote{\url{http://hashpn.space/}} \citep{Parker2016, Bojicic2017}.

A large majority of the faint planetary-nebula nuclei (PNNi) selected for our survey have blue colors, and our HET spectra confirm them as very hot stars. However, in a few cases a redder object lies near the center of the target. These are often chance superpositions of a field star upon the nebula, but sometimes the cool star is in fact physically associated with the PN\null. Two examples from our survey are the rapidly rotating K0~III nucleus of the PN Pa~27, discussed in \citet{Bond_Pa27_2024}; and the dwarf G-type barium central star of the PN K~1-9, presented in Paper~VIII\null. These late-type stars are likely to be binary companions of an optically faint but much hotter true PNN\null. Our HET spectra have also revealed several cases of cataclysmic binary stars which are launching fast winds into space, creating bow-shock nebulae as they interact with the interstellar medium \citep[ISM; see][and references therein]{BondFYVul2025}.

In this paper we investigate two faint nebulae, discovered by amateurs, that host  late-type variable stars. We present deep images showing that the nebulae 
exhibit unusual morphologies. Unlike Pa~27 and K~1-9, discussed in the previous paragraph, these nebulae are almost certainly not ``true'' or classical PNe, in the sense that they are not material ejected during the late evolution of an asymptotic-giant-branch (AGB) star and then ionized by radiation from the star's exposed hot core. We then discuss spectroscopy of both central stars, showing them to be two new members of the class of symbiotic binaries. 

\section{S\MakeLowercase{t}D\MakeLowercase{r} O\MakeLowercase{bjet} 14 and its Variable Central Star, GR C\MakeLowercase{ygni} \label{GR_Cygni} }

\subsection{Discovery of the Nebula}

In the course of selecting targets for our spectroscopic survey, we examined a compilation of candidate PNe discovered by French amateur Xavier Strottner and German amateur Marcel Drechsler.\footnote{A video at \url{https://www.youtube.com/watch?v=71rkTHbeqd8} displays an extraordinary collection of followup deep images of selected faint nebulae discovered by the Strottner-Drechsler team.} These candidates are cataloged at PNe.net with prefixes ``StDr'' and ``StDr Objet.'' Here we focus on \StDr,\footnote{Note that StDr~14 = PN~G047.3+01.2 is a distinct and unrelated likely PN\null. Its central star is also unusual, and will be discussed in a future paper.} which was noted by Strottner and Drechsler on imaging from the Space Telescope Science Institute Digitized Sky Survey\footnote{\url{https://archive.stsci.edu/cgi-bin/dss_form}} (DSS). The PNe.net website presents imagery of the site of this nebula from several sky surveys, but only in the DSS red-bandpass image is it readily seen, as a faint roughly circular nebula with a diameter of about $60''$. In the HASH catalog, \StDr\ is listed as a candidate PN, designated PN\,G080.9$-$06.4 based on its Galactic coordinates. 

\subsection{GR Cyg: Central Star of \StDr}

Color images of \StDr\ at PNe.net show a relatively bright and extremely red star centered within the nebula. Since the nuclei of PNe are almost always faint and blue stars, as discussed in our Introduction, we initially dismissed the star as an unrelated field object---until we realized that it is a known variable star, GR~Cygni. This made the object worth further investigation, in particular the deep imaging and spectroscopy described below.

Table~\ref{tab:GRCyg_DR3data} lists basic data for \GR, taken from \Gaia\/ Data Release~3\footnote{\url{https://vizier.cds.unistra.fr/viz-bin/VizieR-3?-source=I/355/gaiadr3}} (DR3; \citealt{Gaia2016, Gaia2023}). The distance given in the tenth row is from the Bayesian analysis of {\it Gaia\/} astrometric data by \citet{BailerJones2021}. In spite of the star's extremely red color, \GR\  was detected in the near-UV by the {\it Galaxy Evolution Explorer\/} (\GALEX; \citealt{Morrissey2007})\null. The final row in Table~\ref{tab:GRCyg_DR3data} lists the star's near-UV magnitude, quoted from the \GALEX\/ source catalog.\footnote{\url{https://galex.stsci.edu/gr6/?page=mastform}}

\begin{deluxetable}{lc}[b]
\tablecaption{\Gaia\/ DR3 and \GALEX\/ Data for GR Cygni, Central Star of \StDr\ \label{tab:GRCyg_DR3data} }
\tablehead{
\colhead{Parameter}
&\colhead{Value}
}
\decimals
\startdata
RA (J2000) &  21 04 40.078\\
Dec (J2000) & +37 16 41.62 \\
$l$ [deg] & 080.92 \\
$b$  [deg] &  $-$06.47  \\
Parallax [mas] & $0.5950\pm0.0935$ \\
$\mu_\alpha$ [mas\,yr$^{-1}$] & $8.114\pm0.083$ \\
$\mu_\delta$ [mas\,yr$^{-1}$] & $9.553\pm0.097$ \\
$G$ [mag] &  11.88 \\
$G_{\rm BP}-G_{\rm RP}$ [mag] & $3.71$ \\
Distance [pc] & $1799^{+386}_{-265}$ \\
$m_{\rm NUV}$ [AB mag] & $19.43\pm0.08$ \\
\enddata
\end{deluxetable}

The first mention of the spectrum of \GR\ in the literature, to our knowledge, was its inclusion in the \citet{Stephenson1976} {\it General Catalogue of S Stars}. This classification appears to have been based on a low-dispersion objective-prism spectrum. 
\GR\ is a bright source in the near- and mid-IR\null. \citet{Volk1989} examined mid-IR spectra of the star obtained with the slitless low-resolution spectrometer (LRS) on the {\it Infrared Astronomical Satellite\/} (\IRAS)\null. They noted the presence of SiC emission at 11~$\mu$m in the LRS spectrum of \GR, a characteristic of mass-losing carbon stars; \GR\ was singled out as having unusually strong SiC emission for a star that had been classified as of type S in the earlier literature. \citet{Bidelman1991} followed up on this finding by examining a near-IR objective-prism spectrum, and concluded that the S-type classification was incorrect and that the star should be regarded as intermediate between a carbon and S type, i.e., type CS\null. \citet{Kwok1997} reported confirmation of the CS type in a compilation of optical spectral classifications for thousands of \IRAS\/ LRS sources, but without details.

%However, based on examination of infrared objective-prism plates, it was classified as a carbon star by \citet{Alksnis1987}. 

\subsection{Variability of GR Cyg \label{sec:GRCyg_variability} }

A new variable star in Cygnus was among several dozen discovered exactly a century ago by \citet{Ross1926}. He detected variability of the object---subsequently designated \GR---by comparing photographic plates taken about a decade and a half apart using the Bruce Telescope system at Yerkes Observatory (for the main purpose of discovering stars with large proper motions). Ross reported that the star was detected at photographic magnitude 11 on plates from 1909, but was found to be invisible on exposures obtained in 1925. More than eight decades later the Bruce plates were reexamined by \citet{Osborn2012}, and the magnitudes converted to the modern $B$ system. These authors found \GR\ at $B=13.0$ for the 1909 observation, and detected it near the plate limit at about $B=15.8$ at the 1925 epoch.

The variability of \GR\ attracted little attention for more than six decades after Ross's discovery. \citet{Kozyrev1988} presented photometry derived from 145 photographic plates obtained between 1975 and 1982, showing irregular variability over a range of $B$ magnitudes from 11.9 to 15.6. 

With the advent of large-scale all-sky imaging surveys, several groups investigated the variations of \GR. These studies led to its classification as a periodic pulsator, but with a range of derived periods. \citet{Wozniak2004}, based on photometry carried out over about one year by the Northern Sky Variability Survey, classified the star as a Mira variable with a period of 384~days---which is uncertain since the coverage interval was comparable to the period. In the first catalog of variable stars measured by the Asteroid Terrestrial-impact Last Alert System (ATLAS; \citealt{Heinze2018}), \GR\ is again classified as a Mira variable, with a period of 336~days. Here the interval of observations was about two years.  \citet{Chen2020}, based on the first 15~months of data from the Zwicky Transient Facility, classified the star yet again as a Mira pulsator, this time with a period of 456~days. Lastly, \citet{Arnold2020},
using $R$-band data from the Kilodegree Extremely Little Telescope (KELT), obtained over nearly six years, found a period of $459\pm46$~days.

\begin{figure}[b]
\centering
\includegraphics[width=0.47\textwidth]{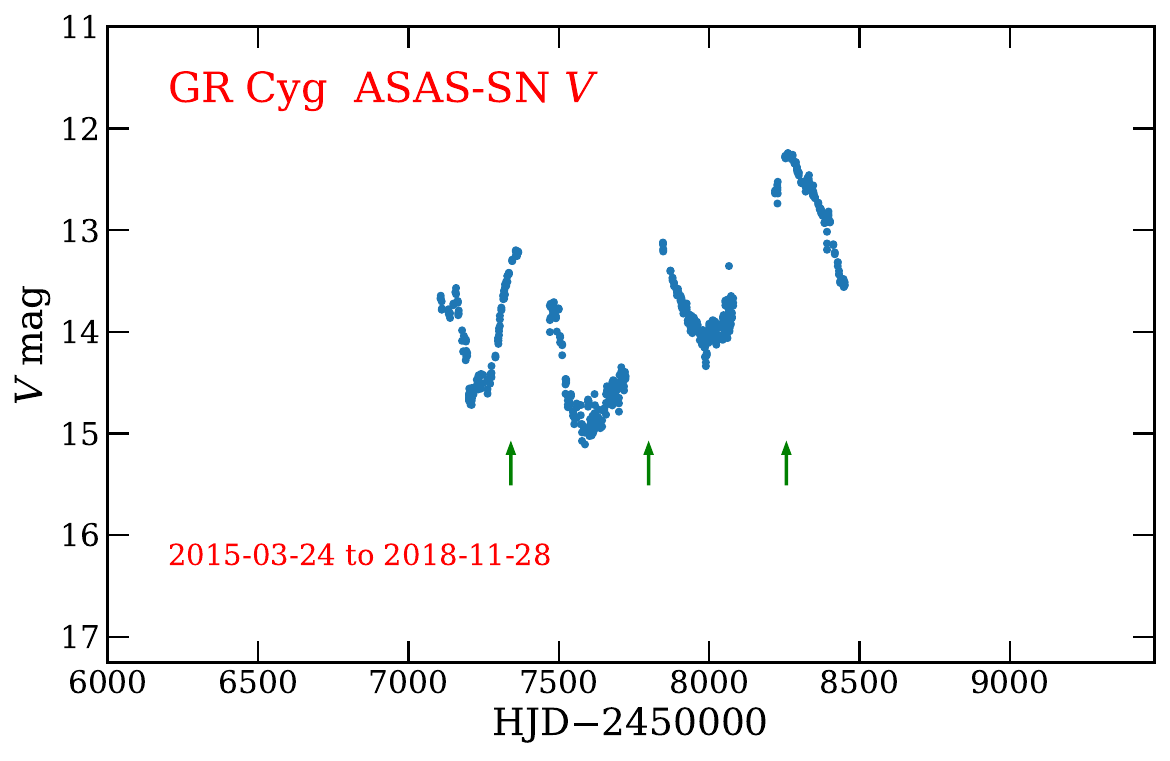}
\includegraphics[width=0.47\textwidth]{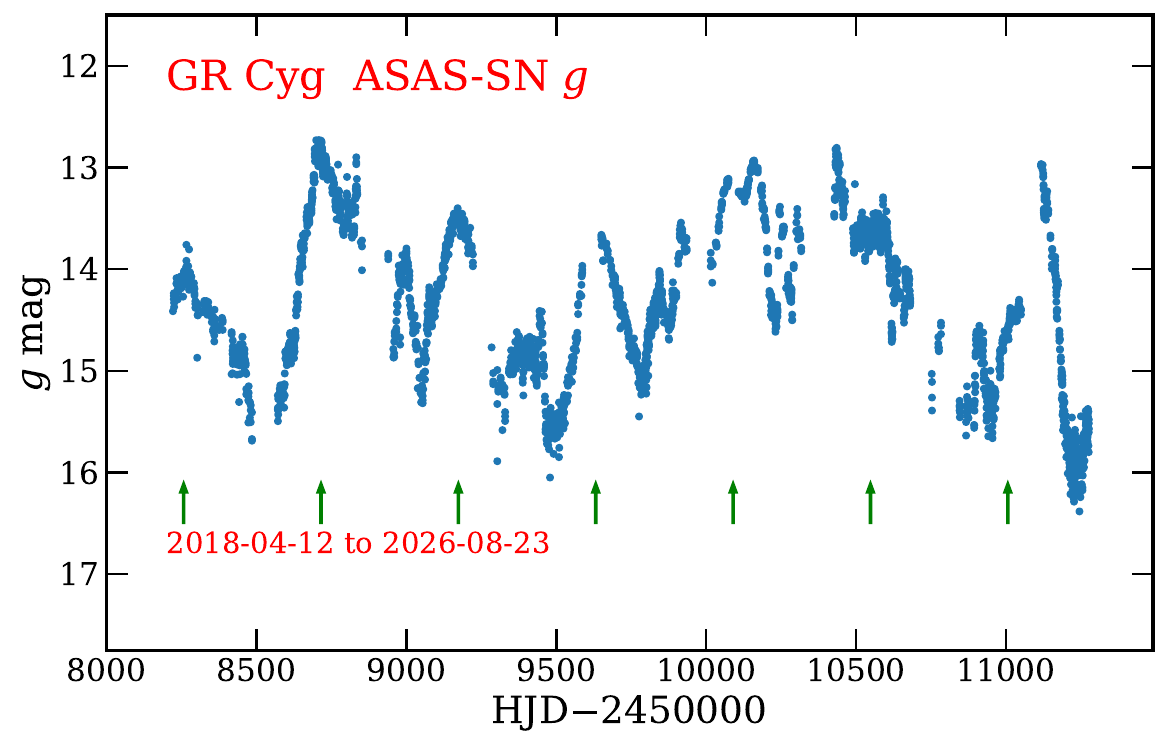}
\caption{
ASAS-SN light curves of \GR\ in $V$(top panel) and $g$ (bottom panel), revealing a mean pulsation period of $\sim$445~days superposed with shorter-term activity. Green arrows are plotted at predicted times of maximum (see text). 
\label{fig:GRCyg_lightcurve}
}
\end{figure}

We downloaded\footnote{From \url{https://asas-sn.osu.edu/}} photometry of \GR\ obtained by the sky patrol of the All-Sky Automated Survey for SuperNovae \citep[ASAS-SN;][]{Shappee2014, Kochanek2017}. The top panel in Figure~\ref{fig:GRCyg_lightcurve} shows the ASAS-SN light curve in the $V$ filter from 2015 March to 2018 October, and the bottom panel plots the light curve in the $g$ band from 2018 April to 2026 August. The variations shown in these panels are not the smooth curves of a classical Mira variable{---a point to which we return in Section~\ref{sec:GRCyg_discussion}.} 
Instead, based on their data, ASAS-SN classifies \GR\ as a semiregular variable (SRV).

For a recent comprehensive review of SRVs and other late-type pulsating variable stars, see \citet{Skowron2026}. A website\footnote{\url{https://ogle.astrouw.edu.pl/atlas/SRVs.html}} from the OGLE collaboration presents examples of SRV light curves that are similar to that of \GR.

We calculated Lomb-Scargle periodograms \citep{Lomb1976, Scargle1982} for the ASAS-SN data; they show broad peaks around a period of $\sim$450~days, for both data sets. 
In the top panel of Figure~\ref{fig:GRCyg_lightcurve}, a fairly smooth variation is seen, with three peaks of a periodic variation superposed on a rising mean level. In the bottom panel the star appears to have become more active, with
considerable variability on short timescales, but the pulsation period is still present over most of this interval. 

%This likely represents the radial pulsation period of the star.

We fitted a linear ephemeris to the times of the four best-defined light-curve peaks in Figure~\ref{fig:GRCyg_lightcurve}, finding $t_{\rm max}={\rm HJD}\,2457340.3+458.2\,E$. The times of maximum predicted from this ephemeris are marked with green arrows in both panels of the figure. Most of the peak times are predicted quite well, but the ephemeris appears to break down for the final predicted time of maximum in the bottom panel.

\begin{figure}[h]
\centering
\includegraphics[width=0.47\textwidth]{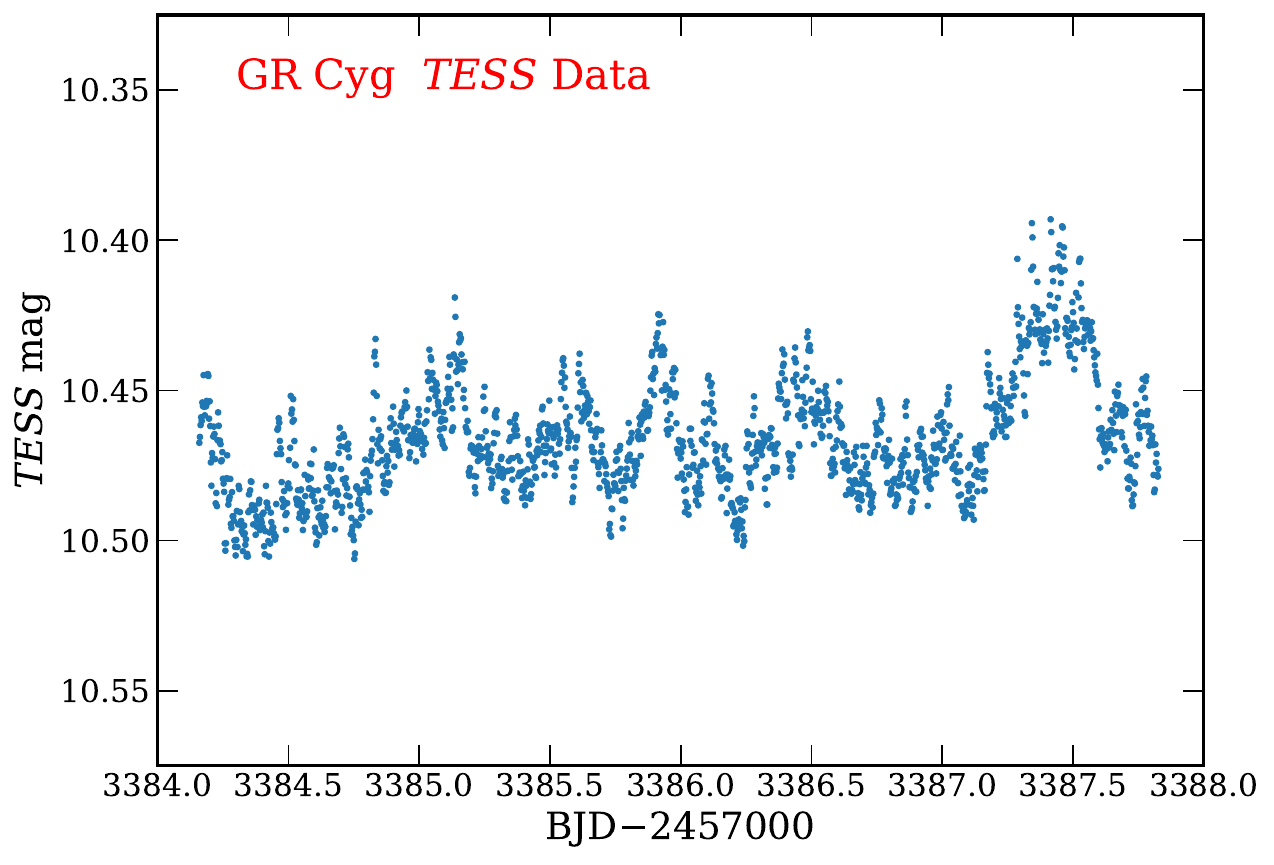}
\caption{
Typical \TESS\/ light curve of \GR, in this case for four days in Sector~76 in 2024 March. Flickering is seen on timescales as short at the observing cadence of 200~s. Uncertainties of individual points are about 0.0012~mag. 
\label{fig:GRCyg_TESS}
}
\end{figure}

Fast-cadence photometry of \GR\ is available from the {\it Transiting Exoplanet Survey Satellite\/} (\TESS; \citealt{Ricker2015}) mission. We obtained \TESS\/ data using the online {\tt TESSExtractor} tool\footnote{\url{https://www.tessextractor.app}} \citep{Brasseur2019, Serna2021}. A representative four-day segment of the \TESS\/ light curve is shown in Figure~\ref{fig:GRCyg_TESS}. Rapid variations (``flickering'') are present, with peak-to-peak amplitudes as large as $\sim$0.05-0.10~mag,\footnote{The true amplitudes are higher, since {\tt TESSExtractor} indicates that 43\% of the measured flux is due to neighboring stars in the crowded field.} and timescales as short as the 200~s observing cadence.

\section{S\MakeLowercase{t}D\MakeLowercase{r} 4 and its Variable Central Star, \D141  \label{StDr_4} }

\subsection{Discovery of the Nebula}

StDr~4 is another low-surface-brightness nebula discovered by
Strottner and Drechsler.\footnote{This nebula should not be confused with StDr Objet~4, an unrelated supernova remnant.} It is included in a list of PNe candidates found by amateurs and publshed by \citet{LeDu2020LAstr}.\footnote{The list is available online at \url{https://vizier.cds.unistra.fr/viz-bin/VizieR?-source=J/other/LAstr/136.58}}  Lying at a low Galactic latitude in Cepheus, StDr~4 was noted in images from the INT Photometric \Ha\ Survey of the Northern Galactic Plane (IPHAS; \citealt{Drew2005}) and from the DSS\null. Its dimensions are given as $3\times2.3$~arcmin. A nebular spectrum obtained by P.~Le\,D\^u, shown at PN.net, indicates a low excitation level, with strong [\nii] emission lines surrounding a weak \Ha, and no obvious emission at [\oiii] $\lambda$5007. [\sii] $\lambda$6716--6731 emission is strong as well. At both PN.net and in the HASH catalog (where it is designated PN G098.4+02.2), StDr~4 is called an ``object of unknown nature.'' Sky-survey images available at the PN.net and HASH sites show a parabolic arc, strongly suggesting a bow-shock morphology for this faint nebula.

\subsection{\D141: Central Star of StDr~4}

Color images of StDr~4 at PN.net reveal a conspicuous, extremely red star lying within the nebula. As in the case of \GR, this discrepancy with the normally blue colors of PNNi prompted us to follow up with deep imaging of the nebula and spectroscopy of the star.

Unlike \GR, the literature on this relatively bright star is surprisingly sparse. \citet{Dolidze1975}, on the basis of an objective-prism spectrum indicating an M8 type with emission lines, included it in a list of nearly 200 probable long-period variable stars. The star is designated ``[D75]~141'' in SIMBAD.\footnote{\url{https://simbad.cds.unistra.fr/simbad}}  We note that \citet{Skiff1997} provided accurate coordinates for many of the stars listed by Dolidze, but had been unable to identify \D141\ itself. However, in a more recent study, Skiff (private communication) used Dolidze's finding chart to confirm an identification of \D141\ with the IR source IRAS\,21398+5533. The only other mention of \D141\ in the literature of which we are aware is its inclusion in a list of periodic variable stars discovered by the Zwicky Transient Facility (ZTF), published by \citet{Chen2020}. These authors classified \D141\ as an SRV with a period of 103.4~days.

In Table~\ref{tab:D141_DR3data} we list astrometric, photometric, and radial-velocity data for \D141, from \Gaia\/ DR3. The distance given in the table is taken from \citet{BailerJones2021}. As the table notes, \GALEX\/ imaged the site of \D141, but did not detect a source at its position. 

\begin{deluxetable}{lc}[t]
\tablecaption{\Gaia\/ DR3 and \GALEX\/ Data for \D141, Central Star of StDr~4 \label{tab:D141_DR3data} }
\tablehead{
\colhead{Parameter}
&\colhead{Value}
}
\decimals
\startdata
RA (J2000) &  21 41 27.870\\
Dec (J2000) & +55 47 29.50 \\
$l$ [deg] & 098.43 \\
$b$  [deg] &  +02.24  \\
Parallax [mas] & $0.5407 \pm0.0653 $ \\
$\mu_\alpha$ [mas\,yr$^{-1}$] & $-5.017 \pm0.072 $ \\
$\mu_\delta$ [mas\,yr$^{-1}$] & $ -5.082\pm0.069 $ \\
$G$ [mag] &  10.56 \\
$G_{\rm BP}-G_{\rm RP}$ [mag] & $4.78$ \\
Radial velocity [$\kms$] & $-43.6 \pm 1.2$ \\
Distance [pc] & $1805^{+219}_{-202} $ \\
$m_{\rm NUV}$ [mag] & (no detection) \\
\enddata
\end{deluxetable}

\subsection{Variability of \D141}

\begin{figure}[b]
\centering
\includegraphics[width=0.47\textwidth]{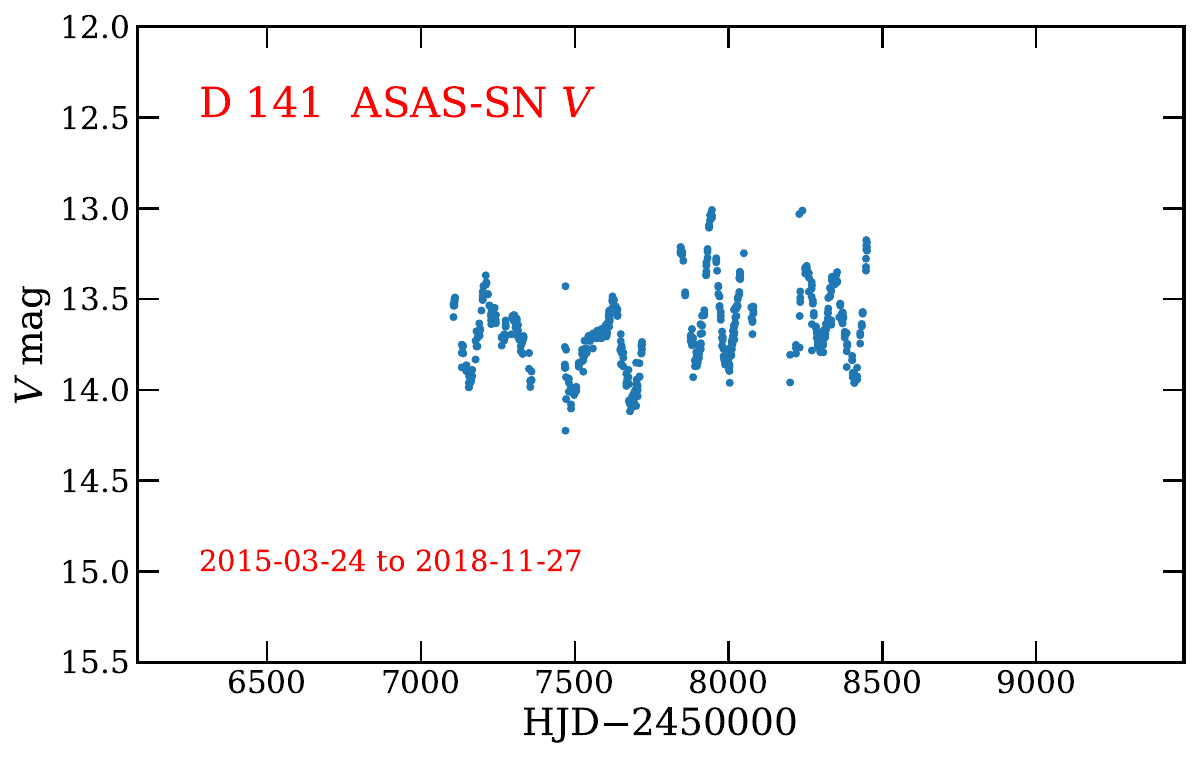}
\includegraphics[width=0.47\textwidth]{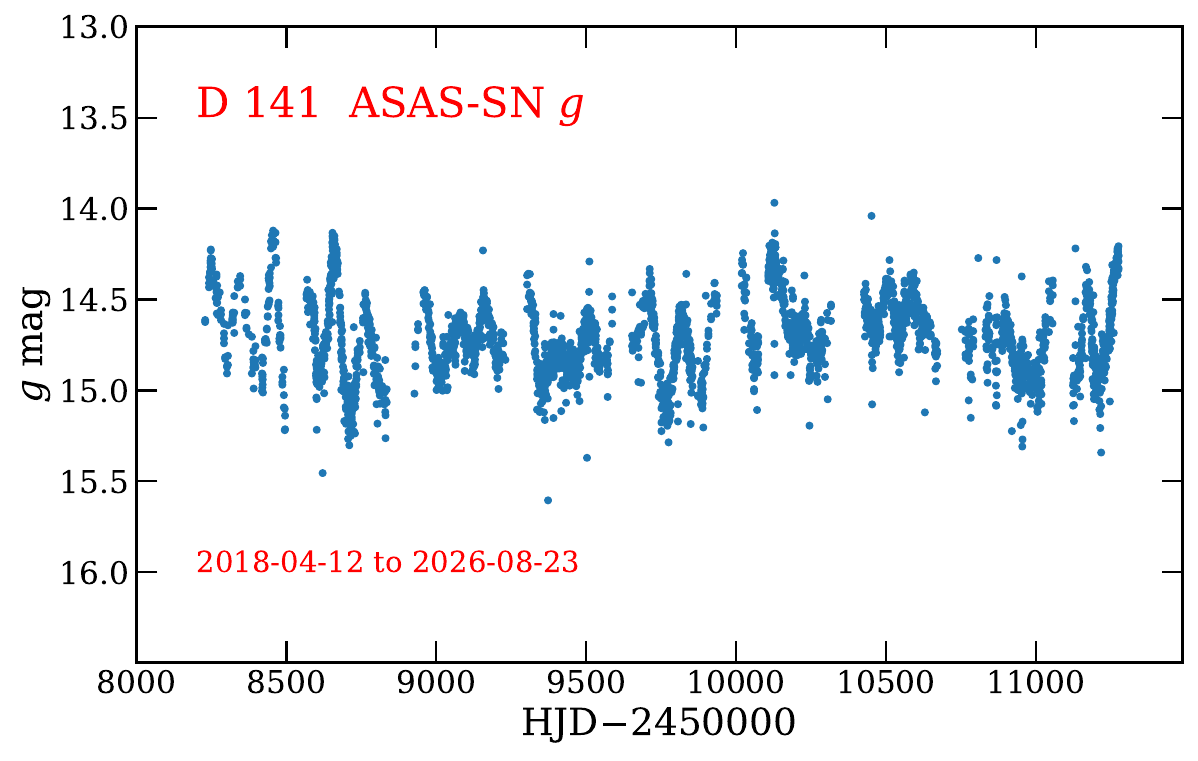}
\caption{
ASAS-SN light curves of \D141\ in $V$ (top panel) and $g$ (bottom panel). The star shows a pulsation period of about 102~days, but with considerable short-term activity.\label{fig:D141_lightcurve}
}
\end{figure}

We downloaded photometry of \D141\ from the ASAS-SN website. The top panel in Figure~\ref{fig:D141_lightcurve} shows the ASAS-SN light curve in the $V$ filter from 2015 March to 2018 October, and the bottom panel plots the light curve in the $g$ band from 2018 April to 2026 August. Both light curves exhibit  variability over a peak-to-peak range of about one magnitude.

A Lomb-Scargle periodogram for these data shows a strong signal at a period of 102.5~days for the $V$-band data, in good agreement with the ZTF result of \citet{Chen2020} mentioned above. The $g$-band behavior is more complex, with the periodogram showing two prominent peaks. The stronger is at a period of 105.1~days, and the weaker is at 97.1~days.  

In summary, as in the case of \GR, the light curves of \D141\ represent typical behavior of an SRV\null. However, the underlying period of \D141\ is considerably shorter than  in \GR.

We examined \TESS\/ data for \D141. The light curves (not shown here) show no rapid flickering such as seen in \GR\ (see Figure~\ref{fig:GRCyg_TESS}). However, they do exhibit low-amplitude solar-like oscillations, typical of late-type red giants. Similar oscillations may be present in \GR, but they are swamped by the star's rapid flickering.

\clearpage

\section{Deep Imaging}

\subsection{S\MakeLowercase{t}D\MakeLowercase{r} O\MakeLowercase{bjet} 14/\GR 
\label{sec:GRCygDeepImaging} }

Following up on our realization that \StDr\ is a nebula of unusual interest, three co-authors of this paper (Carter, Goodhew, and Talbot) obtained deep narrow- and broad-band direct imaging. Observations were made with four telescopes, located at a site in Spain, and in Mississippi and Texas in the USA\null. The telescopes, with apertures of 6 to 13.8~inches, are equipped with low-noise CMOS cameras, allowing large numbers of individual short subexposures to be accumulated over multiple nights and then combined. Table~\ref{tab:StDrObj14_imaging_exposures} gives details of the instrumentation and exposures, carried out in 2025 August and September. A smaller number of additional exposures was obtained in 2026~June. Filters were used that cover narrow-band \Ha\ $\lambda$6562 and [\oiii] $\lambda$5007, and broad-band {\it RGB}\null. We note that the bandpass of the \Ha\ filters has a width of $\sim$50~\AA, so that they also have transmission at the neighboring [\nii] emission lines at 6548 and 6583~\AA\null.

\begin{deluxetable*}{lcccccccc}[h]
\tablecaption{Imaging Exposure Times [s] on \StDr\ \label{tab:StDrObj14_imaging_exposures} }
\tablehead{
\colhead{Observer}
&\colhead{Location\tablenotemark{a}}
&\colhead{Telescope(s)}
&\colhead{Camera\tablenotemark{b}}
&\colhead{\Ha+[\nii]}
&\colhead{[\oiii]}
&\colhead{$R$}
&\colhead{$G$}
&\colhead{$B$}
}
\startdata
Goodhew & (1) & Twin APM 6 in ($f/7.9$) & (1) & $689\times300$ & $85\times300$ & $27\times300$ & $28\times300$ & $29\times300$ \\
Talbot  & (2) & Stellarvue SVX 152T 6 in ($f/8$) & (2) & $59\times1200$ & \dots         & \dots & \dots & \dots \\
Carter \& Talbot & (3) & PlaneWave DR350 13.8 in ($f/3$) & (3) & $28\times480$ & \dots         & \dots & \dots & \dots \\
\noalign{\vskip0.05in}
Total exp.\ [hr] & & & & 80.82 & 7.08 & 2.25 & 2.33 & 2.42 \\
\enddata
\tablenotetext{a}{Telescope locations and observation dates: (1)~Fregenal de la Sierra, Spain, 2025 August 8-12; (2)~Stark Bayou Observatory, Ocean Springs, MS, 2025 August 9-18; (3)~Dark Sky Observatory, Fort Davis, TX, 2025 September 2-3 and 2026 June 9-10.}
\tablenotetext{b}{CMOS cameras and filters: (1)~QHY 268, Astrodon \Ha\ and [\oiii], and Tru-balance $RGB$; (2)~ZWO ASI6200MM Pro, Astrodon \Ha; (3)~ZWO ASI461MM Pro, Chroma \Ha.}
\end{deluxetable*}

\begin{figure*}
\centering
\includegraphics[width=6in]{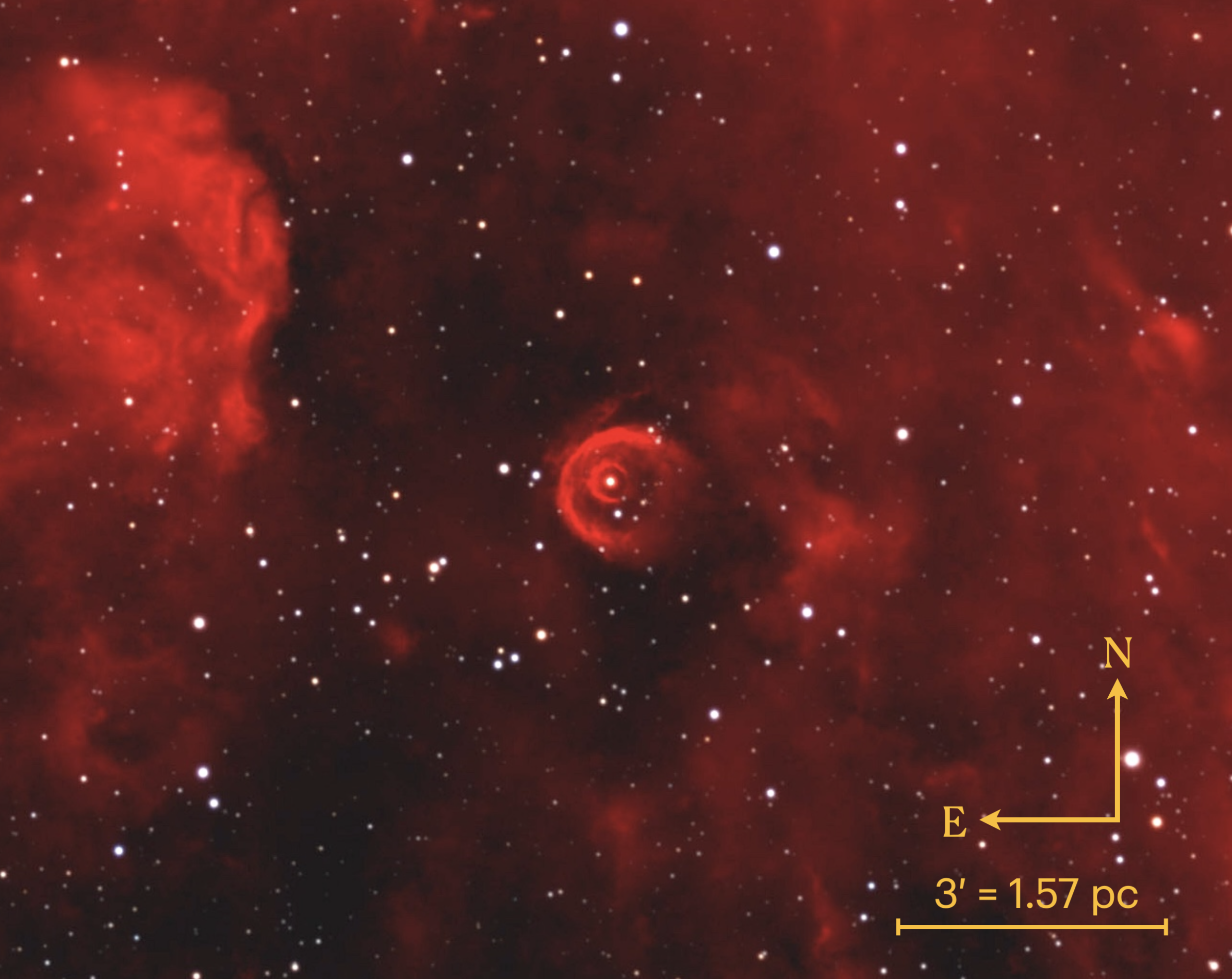}
%\vskip-0.4in
\caption{
Color image of the nebula \StDr\ surrounding the variable star \GR, created from 87.8~hours of exposure time in $RGB$ and \Ha+[\nii] filters, as described in the text. \Ha+[\nii] is assigned to the red channel in this rendition. Orientation and scale of the image are indicated at the lower right, including a conversion to linear scale at the distance of \GR. 
\label{fig:GRCyg_colorimage}
}
\end{figure*}

Co-author Goodhew combined the frames and prepared a color rendition, using software described in Section~5.1 of Paper~VIII\null. The red channel was assigned to the \Ha+[\nii] filter. Since the nebula was not detected in exposures in the [\oiii] filter, this bandpass was not included in the palette. Figure~\ref{fig:GRCyg_colorimage} displays the resulting color image. 

The figure shows that \GR, at its low Galactic latitude, lies in a field with substantial ambient nebulosity. \GR\ itself is seen to be surrounded by two concentric rings. The appearance of the rings appears consistent with an origin in edge-brightened hollow quasi-spherical shells. Our broad-band frames barely show the rings, confirming that they are due to \Ha\ (and\slash or [\nii]) emission from ionized gas.\footnote{Concentric rings are a morphology often seen in dust-scattered light echoes. However, images taken $\sim$9~months apart with the PlaneWave 13.8~in telescope show no expansion of the rings, ruling out a light-echo origin.} The radii of the two rings are about $10\farcs8$ and $29\farcs9$, corresponding to linear radii of 0.09 and 0.26~pc, respectively, at the distance of \GR. 

\begin{figure}
\centering
\includegraphics[width=0.47\textwidth]{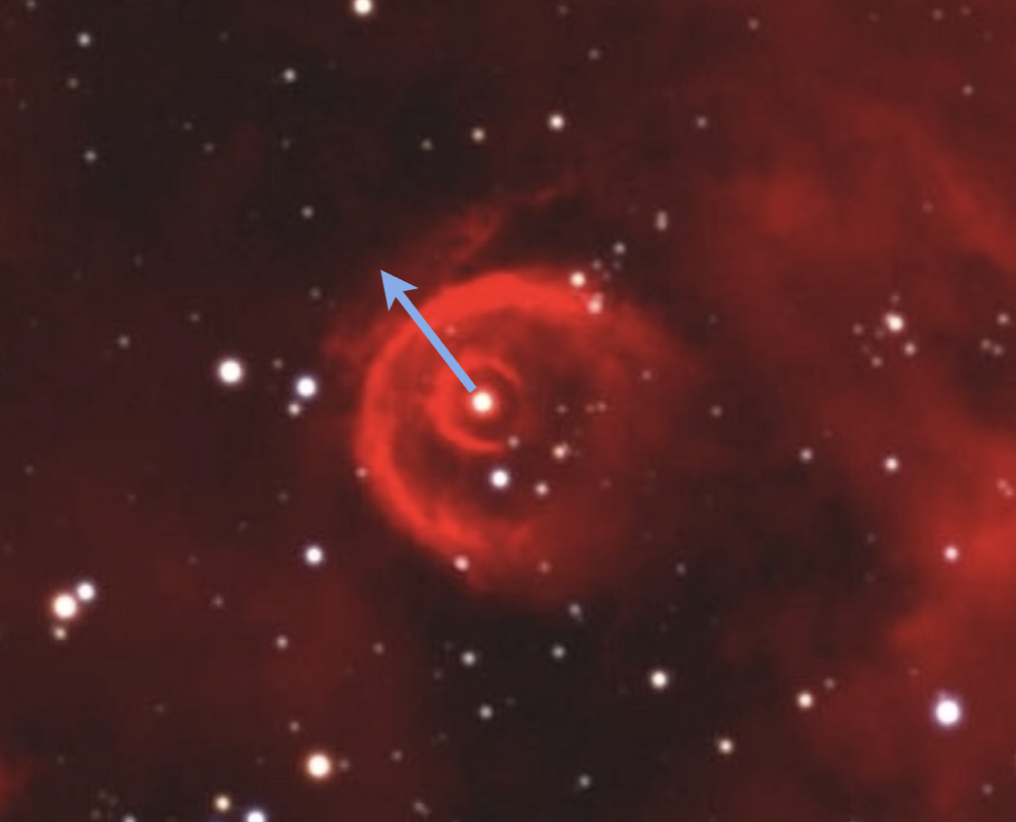}
%\vskip-0.4in
\caption{
Close-up of the \StDr\ nebula. Width of frame is $285''$. The arrow extending from the position of \GR\ is oriented in the direction of the star's proper motion relative to the Galactic-rotation standard of rest at the star's location. See text for details.
\label{fig:GRCyg_zoom}
}
\end{figure}

Figure~\ref{fig:GRCyg_zoom} zooms in on the nebula to show its structure more clearly. Superposed is a blue arrow showing the direction of the central star's proper motion (PM)\null. Here the absolute position angle of the PM of \GR\ derived from the data in Table~\ref{tab:GRCyg_DR3data} has been 
adjusted by $-2\fdg7$, in order to correct for the effect of differential Galactic rotation.\footnote{For details of this correction, see the discussion in \citet{BondLSPeg2025}. To calculate it, we used a {\tt python} code created by S.~del Palacio, along with Oort constants, as referenced in that paper.} The direction depicted is thus made relative to the local standard of rest (LSR) at the distance of the star. The total PM of the star, after this correction, is $16.95\,\masyr$. This corresponds to a high transverse space velocity of \GR, relative to this LSR, of $145\pm27\,\kms$.

This is a highly supersonic velocity, as in the warm interstellar medium the sound speed is about $10\,\kms$. An interaction with the ISM is apparent in Figure~\ref{fig:GRCyg_zoom}, as the two rings are brightest on the leading edge of the motion. On the trailing edge, the rings are considerably fainter, and slightly more distant from the star. Here the outer ring has a diffuse appearance, as the collision with the ISM causes the material to disperse.

\clearpage

\subsection{S\MakeLowercase{t}D\MakeLowercase{r}~4/\D141 
\label{sec:StDr4DeepImaging} }

Co-authors Goodhew and Eklund had imaged StDr~4 in 2022, using their 6- and 14-inch telescopes in Spain. A total of 57.7~hr of exposure time was accumulated in [\sii] $\lambda$6716--6731, \Ha+[\nii], and [\oiii] $\lambda$5007 narrow-band, and $RGB$ broad-band filters.  A color rendition of these data was posted online.\footnote{At \url{https://www.imagingdeepspace.com/stdr-4.html}} However, co-author Bond had been unaware of this imagery when the central star was added to the \hbox{LRS2-B} spectroscopic program in 2025. Later in 2025, based on our newly aroused interest in StDr~4, Goodhew obtained additional exposure time in \Ha+[\nii] with his telescopes, and Talbot also accumulated exposures in the same bandpass with his 3.5-inch telescope in Mississippi. These new data more than double the exposure time on the target to a total of 145.7~hr. Details of all of the exposures from both 2022 and 2025 are given in Table~\ref{tab:StDr4_imaging_exposures}.

\begin{deluxetable*}{lcccccccccc}[h]
\tablecaption{Imaging Exposure Times [s] on StDr~4 
\label{tab:StDr4_imaging_exposures} }
\tablehead{
\colhead{Observer}
&\colhead{Location\tablenotemark{a}}
&\colhead{Telescope(s)}
&\colhead{Camera\tablenotemark{b}}
&\colhead{[\sii]}
&\colhead{\Ha+[\nii]}
&\colhead{[\oiii]}
&\colhead{$R$}
&\colhead{$G$}
&\colhead{$B$}
&\colhead{$RGB$}
}
\startdata
Goodhew & (1) & Twin APM 6 in ($f/7.9$) & (1) & \dots & $787\times300,$ & \dots & $35\times300$ & $31\times300$ & $31\times300$ & \dots \\
  & &  &  &  & $70\times900$ &  &  &  &  \\
Goodhew & (1) & Celestron 14 in ($f/1.9$) & (1) & \dots & $79\times300$ & \dots & \dots & \dots & \dots & \dots\\
Talbot  & (2) & Stellarvue SV80S 3.5 in ($f/\mathbf{6}$) & (2) & \dots & $108\times600$ & \dots         & \dots & \dots & \dots & $114\times120$ \\
Eklund & (3) & Celestron 14 in ($f/1.9$)  & (3) & $41\times600$ & \dots &  $116\times600$        & \dots & \dots & \dots & \dots \\
\noalign{\vskip0.05in}
Total exp.\ [hr] & & & & 6.83 & 107.66 & 19.33 & 2.92 & 2.58 & 2.58 & 3.80 \\
\enddata
\tablenotetext{a}{Telescope locations and observation dates: (1)~Fregenal de la Sierra, Spain, 2022 November~21--December 17 and 2025 July~19--August~25;  (2)~Stark Bayou Observatory, Ocean Springs, MS, 2025 July~21--August~1; (3)~Fregenal de la Sierra, Spain, 2022 November~27.}
\tablenotetext{b}{CMOS cameras and filters: (1)~QHY 268, Astrodon \Ha\ (6~in), Baader \Ha\ (14 in), and Tru-balance $RGB$; (2)~ASI 2600 MC pro color, Andtlia ALP-T \Ha/[\oiii] dual band, ``$RGB$'' =  no filter; (3)~ASI 6200 MM pro, Chroma [\sii] and Baader [\oiii].}
\end{deluxetable*}

\begin{figure*}
\centering
\includegraphics[width=6in]{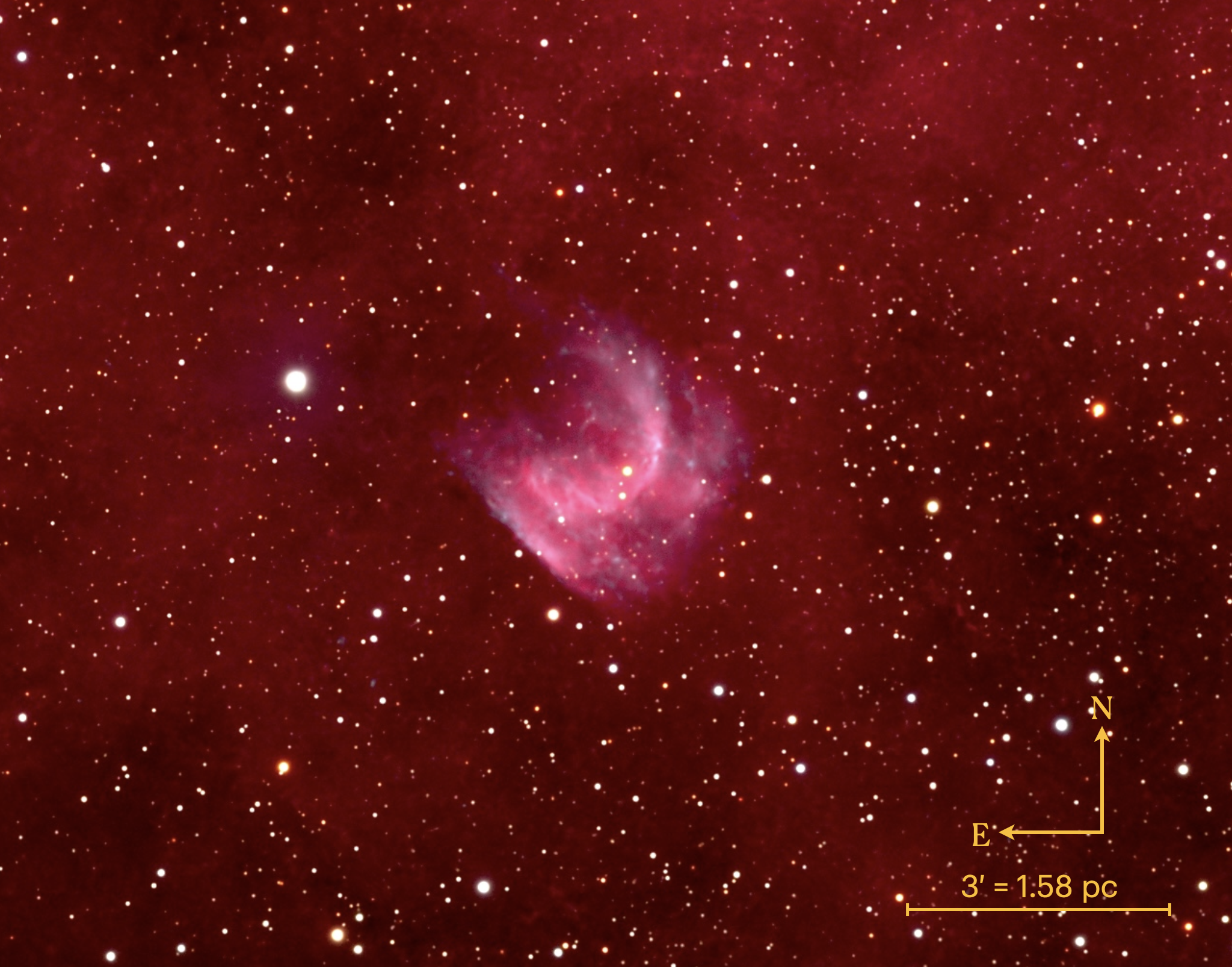}
%\vskip-0.4in
\caption{
Color image of StDr~4 surrounding the variable star \D141, created from exposures in $RGB$, \Ha+[\nii], and [\sii] filters, as described in the text. In this rendition \Ha+[\nii] is assigned to the red channel, and [\sii] to the blue channel. [\oiii] is very faint and is not included. Orientation and scale of the image are indicated at the lower right, including a conversion to linear scale at the distance of \D141. 
\label{fig:StDr4_colorimage}
}
\end{figure*}

Figure~\ref{fig:StDr4_colorimage} presents a color image of StDr~4, prepared by Talbot from the $RGB$, \Ha+[\nii], and [\sii] frames. The processing was entirely with {\tt PixInsight}\footnote{\url{https://pixinsight.com}} software. In order to distinguish between the \Ha+[\nii] and [\sii] frames, which have similar wavelengths, the latter were assigned to the blue channel in this rendition. The nebula is barely detected in [\oiii], so that filter is omitted.

Our deep imagery reveals a bow-shock morphology for the StDr~4 nebula, in the form of a parabolic arc lying to the west and southwest of the star. A bow shock is the result of a fast stellar wind from the central star colliding with the slower surrounding ISM\null. For a recent discussion of astrophysical bow shocks see \citet{Ilkiewicz2026}.

\begin{figure}
\centering
\includegraphics[width=0.47\textwidth]{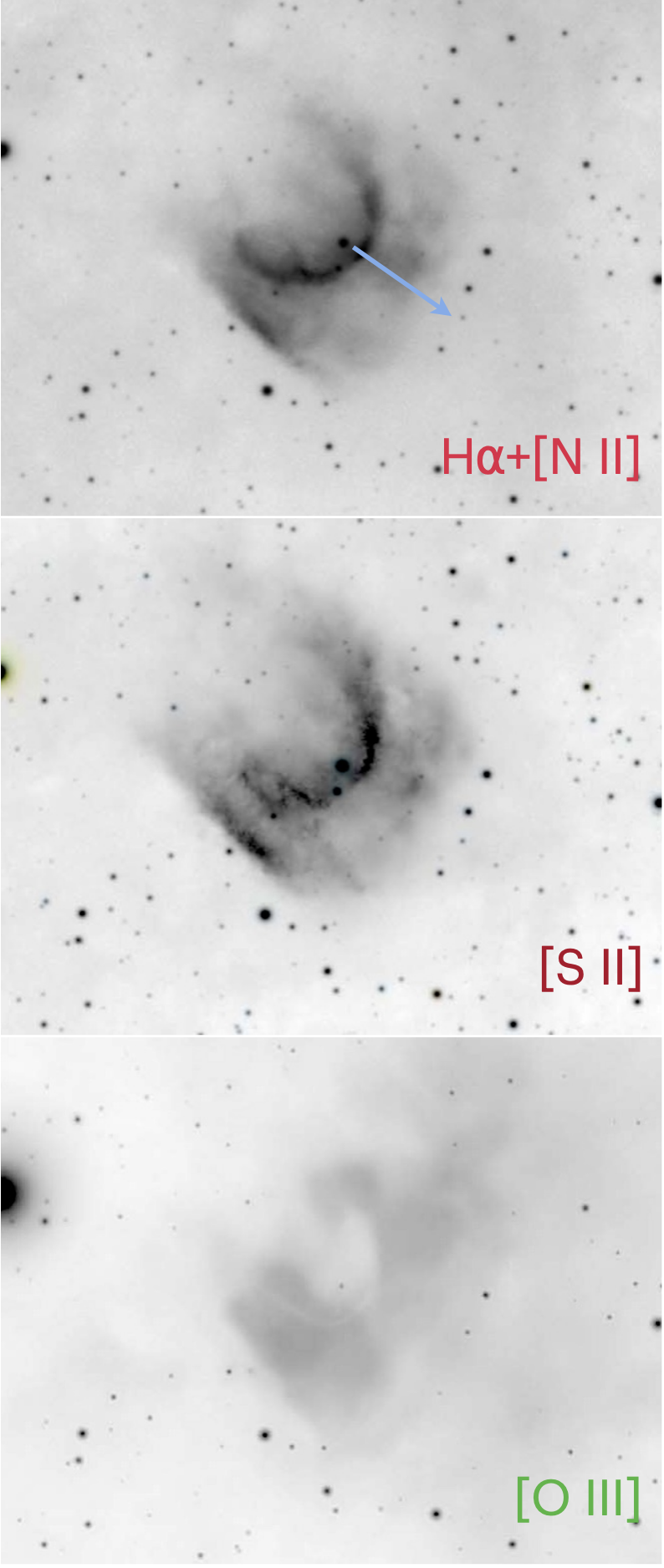}
%\vskip-0.4in
\caption{Narrow-band images of StDr~4 in \Ha\ (top), [\sii] (middle), and [\oiii] (bottom). Width of the frames is $425''$. A blue arrow in the top frame shows the direction of the proper motion of \D141\ relative to the standard of rest at its location. 
\label{fig:StDr4_BandW}
}
\end{figure}

In Figure~\ref{fig:StDr4_BandW} we present images of StDr~4 in three individual narrow-band filters. The top panel shows the nebula in \Ha+[\nii], and the middle panel shows it in [\sii]. These panels show that [\sii] is conspicuously bright, consistent with collisional excitation in the bow-shock interaction. The bottom panel shows [\oiii], which is only weakly detected in this low-excitation nebula. From this we anticipate that the nebulosity in the top panel is due primarily to [\nii] rather than \Ha.

In the top panel of Figure~\ref{fig:StDr4_BandW} a blue arrow shows the direction of \D141's PM\null. The direction of the motion is fully consistent with our interpretation of the parabolic arc as a bow shock. As in the case of \GR\ above, the absolute position angle of the PM has been adjusted, in this case by $+10\fdg4$, in order to correct for the effect of differential Galactic rotation.  The total PM of the star, after this correction, is $2.67\,\masyr$. This corresponds to a transverse space velocity of the star, relative to the LSR at its location, of $22.7\pm1.2\,\kms$. Including the radial velocity from \Gaia\/ DR3, given in Table~\ref{tab:D141_DR3data}, the total space velocity relative to the LSR is $37.3\pm1.1\,\kms$

The StDr~4 bow shock is embedded in a larger, patchy diffuse nebula. The overall morphology of StDr~4 is remarkably similar to that of the nebula Abell~35; see, for example, the deep imagery of Abell~35 obtained by co-author Talbot.\footnote{Available at \url{https://app.astrobin.com/i/rmnn01}} The optical spectrum of the central star of Abell~35 is dominated by a late-type star, which is accompanied by a UV-bright hot white-dwarf (WD) companion. Although originally considered to be a PN \citep[e.g.,][]{Abell1966}, Abell~35 has a shape that is very atypical of PNe. Instead, Abell~35 is now interpreted as the result of the passage of the binary system through a dense region of the ISM; for example, see \citet{Ziegler2012} and references therein. UV radiation from the WD component creates an ionized Str\"omgren zone, within which there is a bow shock excited by the collision of a wind from the system with the ISM\null. StDr~4, with its very similar morphology, likely has a similar origin.

%, with the ionizing flux and fast wind arising from a faint, hot companion accreting from the stellar wind of the late-type optical central star.

\clearpage

\section{Spectroscopy of \GR\ \label{sec:GRCyg_spectroscopy} }

\subsection{Hobby-Eberly LRS2-B Observations \label{sec:HET_obs_of_GRCyg} }

Based on the remarkable morphology of \StDr\ revealed by our imaging (see Section~\ref{sec:GRCygDeepImaging}), we added its central star, \GR, to the target list for our HET\slash LRS2-B spectroscopic survey, described in the Introduction. HET observations are carried out by on-site astronomers in a queue-scheduling mode \citep{Shetrone2007PASP}.
LRS2-B is an integral-field-unit (IFU) spectrograph. It uses 280 $0\farcs6$-diameter lenslet-coupled fibers covering a $12''\times 6''$ field of view (FOV), which feed two spectrograph units via a dichroic beamsplitter.  The ``UV'' channel of LRS2-B covers the wavelength range 3640--4645~\AA\ at resolving power 1910, while the ``Orange'' channel covers 4635--6950~\AA\ at resolving power 1140. 

Data reduction and absolute calibration is carried out by co-author Zeimann, using the \texttt{Panacea}\footnote{\url{https://github.com/grzeimann/Panacea}} and \texttt{LRS2Multi}\footnote{\url{https://github.com/grzeimann/LRS2Multi}} packages. Full details of the LRS2-B spectrograph and general data-reduction procedures are given by \citet{Chonis2016}, and in \citet{Bond2023a}, respectively.  

For each individual exposure, the spectrum of the target star is extracted using a $2''$-radius circular aperture. The local background is estimated from an annular region spanning radii of 3--$5''$; its spectrum is subtracted from the stellar spectrum, removing both night-sky emission and continuum, and any diffuse nebular emission surrounding the star. 

%The use of IFU spectroscopy allows this background to be modeled locally and contemporaneously within the same exposure, which is particularly advantageous for faint PNe with spatially varying surface brightness.

Individual exposures are combined using inverse-variance weighting. Prior to combination, each exposure is scaled according to its signal-to-noise ratio measured near 5100~\AA, ensuring that frames obtained under better observing conditions contribute preferentially to the final spectrum. The combined spectrum is then resampled to a spacing of 0.7~\AA. 

Table~\ref{tab:GRCygLRSexposures} presents an observing log for our HET\slash LRS2-B exposures on \GR, as well as for the Asiago observations described below. Our initial short-exposure spectra were obtained in 2025 August and September.  The LRS2-B IFU makes it possible to create synthetic narrow-band images of the target in the $12''\times 6''$ FOV, as described for example by \citet{BondAbell572024}. We exploited this capability by obtaining a series of longer-exposure observations in 2026 April and May, as listed in Table~\ref{tab:GRCygLRSexposures} and discussed in detail below (Section~\ref{sec:GRCyg_synthetic_imaging}).

%They revealed a remarkable late-type spectrum, superposed by emission lines of the Balmer series, [\oii], [\oiii], \feii, and other species.

\begin{figure*}
\centering
\includegraphics[width=6in]{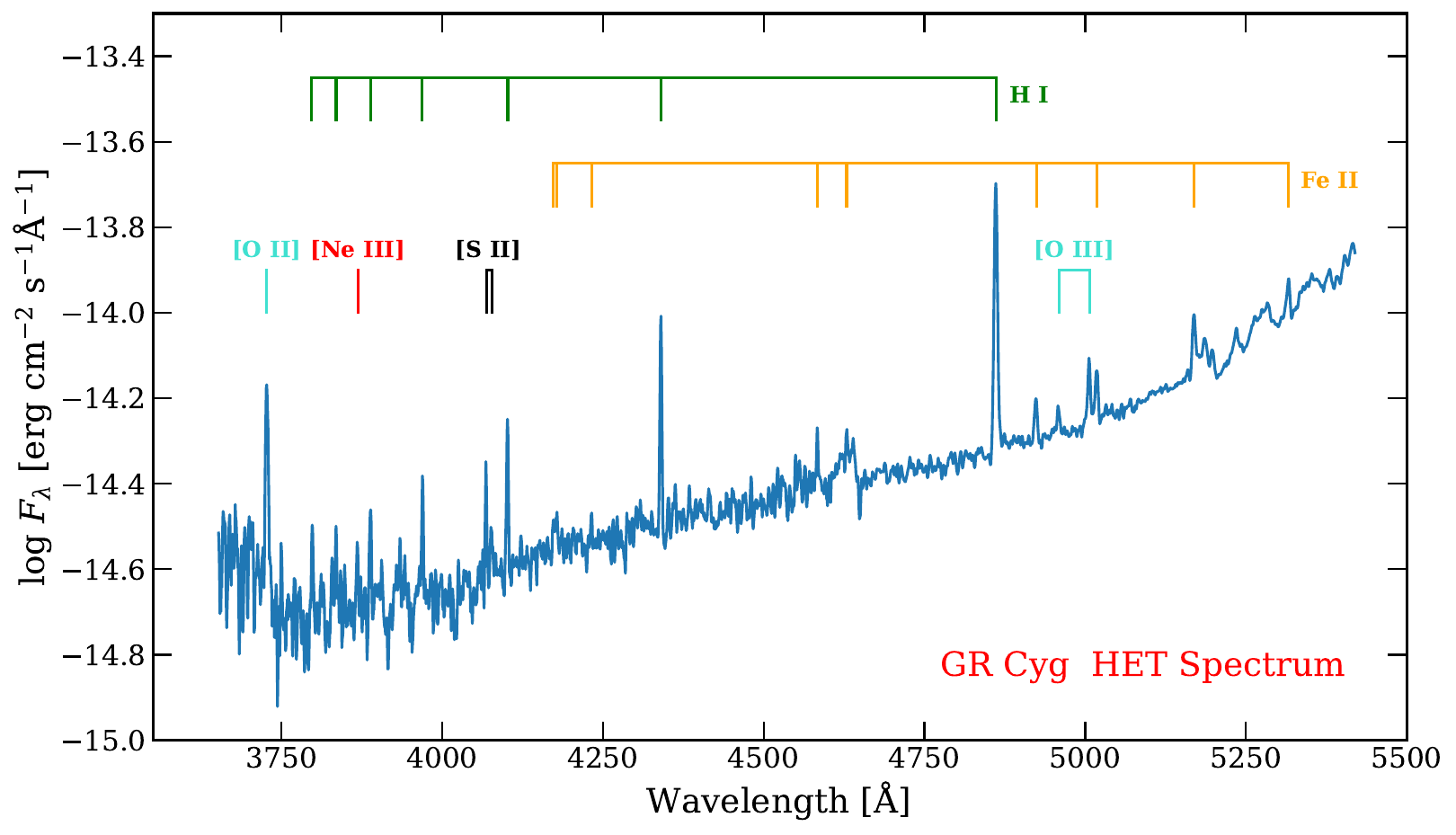}
%\vskip-0.4in
\caption{
Spectrum of \GR\ obtained with the HET LRS2-B spectrograph. Emission lines of the Balmer series, \feii, [\oii], [\oiii], [\neiii], and [\sii] are marked. \label{fig:GRCyg_HETspectrum}
}
\end{figure*}

Figure~\ref{fig:GRCyg_HETspectrum} presents the HET spectrum of \GR\ from 3640 to 5400~\AA, derived by averaging the two exposures from 2025 August and September. A rich emission-line spectrum is seen, superposed on a red continuum. The strongest emission lines are labeled in the figure. The ionization level of these lines is relatively low, showing strong [\oii], [\neiii], and [\sii], along with numerous lines of \feii, as well as the Balmer series. The lines of [\oiii] at 4959--5007~\AA\ are relatively weak. There is a turn-up in the continuum level at the short-wavelength end of the spectrum.

\begin{deluxetable}{lcc}[h]
\tablecaption{Log of HET and Asiago Spectroscopic Obserations of \GR\
\label{tab:GRCygLRSexposures} }
\tablehead{
\colhead{Telescope}
&\colhead{Date}
&\colhead{Exposure}\\
\colhead{}
&\colhead{[YYYY-MM-DD]}
&\colhead{[s]}}
\decimals
\startdata
HET & 2025-08-22 & $36   $	    \\
HET & 2025-09-11 & $38   $	    \\
Asiago 1.22 m & 2026-01-19 & 2400 \\
Asiago 1.82 m & 2026-02-01 & 900 \\
HET & 2026-04-23 & $3\times307$ \\      
HET & 2026-04-26 & $3\times307$ \\      
HET & 2026-05-07 & $3\times307$ \\      
HET & 2026-05-10 & $3\times307$ \\      
HET & 2026-05-12 & $3\times307$ \\      
HET & 2026-05-13 & $3\times307$ \\
\enddata
\end{deluxetable}

\subsection{Asiago Observations}

Co-author Munari obtained follow-up spectra of \GR\ at the Asiago Astrophysical Observatory in 2026 January and February, as well as a photometric observation, just before the star's solar conjunction. Details of the spectroscopic observations are included in Table~\ref{tab:GRCygLRSexposures}. The observation on 2026 February~1 was obtained with the
Asiago 1.82~m telescope and its REOSC-Echelle spectrograph (resolving power $R\simeq22,000$,
wavelength range 3550--7110~\AA).
The Echelle spectrum will be investigated
in a forthcoming paper giving further details of GR~Cyg, and we focus here on the lower-resolution spectrum obtained with the 1.22~m telescope

\subsubsection{Low-Resolution Spectroscopy}

A low-resolution spectrum  of \GR\ was obtained on 2026 January~19.735, with the Asiago
1.22~m telescope equipped with its Boller \& Chivens spectrograph, covering the
4230--6650~\AA\ range at 1.18~\AA\,pix$^{-1}$.  A 600~line\,mm$^{-1}$ grating blazed at
5000~\AA\ was used, and the long slit was set to a width of $1\farcs8$ and
oriented east-west.  

Nearby spectrophotometric standards were observed
immediately before and after the target star, to ensure an accurate absolute
flux calibration.  Data reduction was performed in {\tt IRAF},\footnote{NOIRLab IRAF is distributed by the Community Science and Data Center at NSF NOIRLab, which is managed by the Association of Universities for Research in Astronomy (AURA) under a cooperative agreement with the U.S. National Science Foundation. See \citet{Tody1986} and \citet{Fitzpatrick2025}.} following 
standard procedures; these include
bias, dark, and flat-field corrections, sky subtraction (with the sky median
computed on both sides of the central star and in the region between the two
nebular rings shown in Figure~\ref{fig:GRCyg_zoom}), and wavelength calibration.   

%The zero-point of the flux scale was subsequently refined by comparison with simultaneous $BVRI$ photometry obtained with the Asiago 67/92~cm Schmidt telescope.  

% I don't know why, but citing this reference confuses Overleaf!?
%standard procedures detailed by \citet{Zwitter2000}; these include

%The resulting spectrum is presented in Figure~\ref{fig:GRCyg_Asiagospectrum}. 

%The photometric observations werecalibrated to the \citet{1992AJ....104..340L} standard system via color equations solved on field stars around GR Cyg, selected from the APASS DR8sky survey \citep{2014CoSka..43..518H}.  This yielded magnitudes of \GR\ at the time of observation of $B=16.152, V=12.081, R=9.909$, and $I=8.315$ (with uncertainties of0.014, 0.006, 0.004, and 0.003~mag, respectively, which quadratically combinethe Poissonian component and the transformation errors from the color equations).

\subsubsection{$\mathrm{BVRI}$ Photometry}

Photometric $BVRI$ measurements of GR~Cyg (and of [D75]~141, discussed below) were obtained
on the same nights as their low-resolution spectroscopic observations,
primarily to refine the absolute flux calibrations of the 1.22~m spectra.  For this
purpose, we used the Asiago 67/92~cm Schmidt telescope, equipped with standard
photometric filters and a large-format CCD camera covering a $1^\circ \times 1^\circ$
FOV\null.  The photometric observations were calibrated
to the \citet{1992AJ....104..340L} standard system via color equations
solved on field stars surrounding both targets, selected from the APASS DR8
sky survey \citep{2014CoSka..43..518H}.  The resulting $BVRI$ magnitudes and
their associated uncertainties are listed in Table~\ref{tab:photometry}.  The quoted errors
represent the total error budget, computed by combining in quadrature
the Poissonian noise and the uncertainty from the transformation to the
standard system. \GR\ is extremely red at visible wavelengths, with $B-V=4.07$.

\begin{deluxetable*}{lccccc}[h]
\tablecaption{Asiago Schmidt Photometry of \GR\ and \D141 \label{tab:photometry} }
\tablehead{
\colhead{Star}
&\colhead{Date}
&\colhead{$B$}
&\colhead{$V$}
&\colhead{$R$}
&\colhead{$I$}
}
\startdata
GR Cyg     & 2026-01-19.726 & $16.152 \pm 0.014$ & $12.081 \pm 0.006$ & $ 9.909 \pm 0.004$ & $8.315 \pm 0.003$ \\  
\D141  & 2026-01-20.861 & $15.782 \pm 0.012$ & $13.434 \pm 0.005$ & $11.562 \pm 0.006$ & $9.385 \pm 0.008$ \\ 
\enddata
\end{deluxetable*}

\subsection{Spectral Classification of \GR \label{sec:GRCyg_spectral_classification} }

\begin{figure*}
\centering
\includegraphics[width=6in]{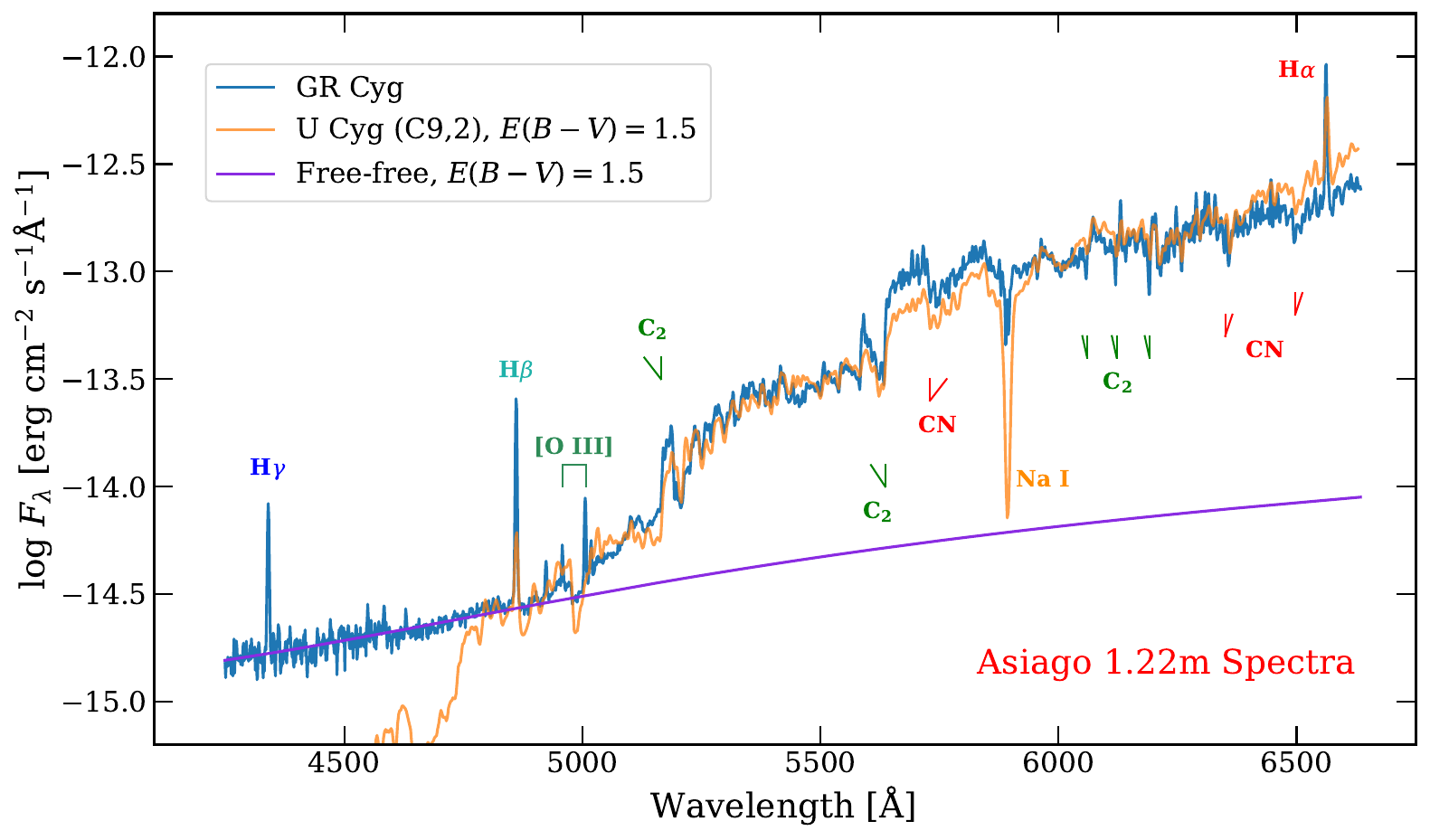}
\caption{
Asiago 1.22~m spectrum of \GR\ (blue line) compared with that of the C9,2 carbon star U~Cyg (orange line). A reddening of $E(B-V)=1.5$ has been applied to U~Cyg. C$_2$ and CN absorption bands and the \nai\ absorption doublet are marked, along with emission lines of the Balmer series and [\oiii]. Note the strong blue excess of \GR\ at the short-wavelength end. The purple curve plots a free-free emission spectrum from a 10,000~K plasma, with the same reddening applied as for U~Cyg, and fitted to the blue flux. See text for discussion.
\label{fig:GRCyg_Asiagospectrum}
}
\end{figure*}

In Figure~\ref{fig:GRCyg_Asiagospectrum}, the Asiago 1.22~m low-resolution spectrum of GR~Cyg is plotted as a blue line. Several strong molecular bands of C$_2$ and CN are marked, along with emission lines of the Balmer series and [\oiii]\null. Superposed as an orange line is a spectrum of the cool
carbon star U~Cygni (HD~193680), which has a spectral type of C9,2 following the classification of \citet{Yamashita1975} and
\citet{1979MNRAS.186..837C}.  U~Cyg provided the closest match to GR~Cyg
among the numerous carbon stars that have been observed with the Asiago 1.22~m spectrograph. In particular, U~Cyg represents a significantly better fit to \GR\ than any S- or CS-type star, since both stars lack the ZrO absorption bands that define the S spectral type.  We note that U~Cyg is a classical Mira variable star with a period of 465~days, a remarkably similar pulsation period to that of the SRV \GR\ (see Section~\ref{sec:GRCyg_variability}). Its spectrum shows Balmer emission lines, as does \GR. These are typically seen in the spectra of pulsating late-type variable stars; however, U~Cyg lacks the forbidden emission lines that are strikingly present in \GR.

The spectrum of GR~Cyg, longward of about 4800~\AA, is extremely red; for U~Cyg to achieve a satisfactory match to the spectral-energy distribution (SED),
it was necessary to artificially redden its spectrum by an additional $E(B-V) =
1.5$, over and above its (small) intrinsic reddening.  The remaining minor discrepancies between the two spectra in the fit of Figure~\ref{fig:GRCyg_Asiagospectrum} above $\sim$4800~\AA\ can be ascribed to several factors: (1)~in order to redden U~Cyg, we adopted the standard $R_V = 3.1$ interstellar
reddening law of \citet{1999PASP..111...63F}, but the actual extinction law
may differ, especially if a significant fraction of the reddening affecting
GR~Cyg is circumstellar in origin;
(2)~the spectra of the coolest giants are highly sensitive to
the precise chemical and even isotopic composition of their upper atmospheres,
where molecules form.  The composition is governed by the evolutionary
history and details of the third dredge-up episodes between thermal
pulses in AGB stars, to the extent that no two carbon stars appear to exhibit precisely identical
spectra \citep[e.g.,][]{1987clst.book.....J}; and (3)~both GR~Cyg and U~Cyg are
pulsating variables, whose spectra vary over their pulsation cycles.

\subsection{\GR: A Newly Discovered Symbiotic Binary \label{sec:GRCyg_new_symbiotic} }

Below about 4800~\AA, where the reddened SED of U~Cyg fades rapidly, the spectrum of \GR\ is dominated by a much hotter
continuum.  To highlight its presence, we fitted this component in Figure~\ref{fig:GRCyg_Asiagospectrum}
with free-free emission at an electron temperature $T_e = 10,000$~K,
reddened by $E(B-V) = 1.5$, and plotted as a purple line.  Along with this hot continuum, the presence of
various strong emission lines, and the detection of the star in the NUV by \GALEX, all support an interacting-binary nature for GR~Cyg, in which the
carbon giant is orbited by a much hotter but optically faint companion.  The free-free emission
could originate from an accretion disk around the hot companion, from the
fraction of the carbon giant's stellar wind ionized by the hot component, or from a
combination of both mechanisms. Because the numerous nebular forbidden emission lines
visible in the HET spectrum (Figure~\ref{fig:GRCyg_HETspectrum}) cannot form in the dense environment
of an accretion disk, they must originate within the ionized wind of the red giant.

A cool red giant interacting with an accreting hot companion defines a symbiotic
binary {(e.g., \citealt{Mukai2016}).} The spectra presented in Figures~\ref{fig:GRCyg_HETspectrum} and \ref{fig:GRCyg_Asiagospectrum} thus establish GR~Cyg as a
newly discovered symbiotic star.  Catalogs of symbiotic stars have been
compiled by several authors \citep[among others,][]{1984PASA....5..369A,
2000A&AS..146..407B, 2019ApJS..240...21A, Merc2026}.  

Symbiotic
binaries harboring a carbon giant are extremely rare in the Galaxy (only about 3\% of the known objects; see \citealt{MercCarbonSymbiotic2025}), where the vast
majority host M-type giants.  Conversely, carbon giants dominate symbiotics in the Magellanic
Clouds, where the ambient metallicity is well below the Galactic value; this is likely due to
low metallicity boosting the efficiency of the third dredge-up in bringing
carbon to the stellar surface in AGB stars.  The high space velocity of \GR\ (Section~\ref{sec:GRCygDeepImaging}) and the weakness of its \nai\ D lines (Figure~\ref{fig:GRCyg_Asiagospectrum})  are consistent with the star's membership in an old metal-poor population.

Symbiotic stars can generally be divided into
two major branches \citep[see the review by][]{2019arXiv190901389M}: the ``burning type,''  in which the accreted material nuclearly burns on the surface of the WD; and the ``accreting-only''
type, in which the accreted hydrogen-rich material accumulates quietly on the surface of the WD without undergoing nuclear fusion. The burning-type symbiotics are spectroscopically conspicuous and constitute the majority of known symbiotic binaries. { They generally show high-excitation emission lines, such as \heii\ $\lambda$4686, which were taken as the defining characteristic of symbiotic stars in earlier literature \citep[e.g.,][]{1984PASA....5..369A}.} However, modern surveys are revealing increasing numbers of accreting-only systems.
The low luminosity of the blue continuum, the moderate intensity of the
emission lines, and their low ionization state all suggest that GR~Cyg is an
accreting-only symbiotic binary.

{

One hallmark of the burning-type symbiotics is the presence of emission features at 6825 and 7088~\AA\null. These are due to Raman scattering of the UV emission doublet of \ovi\ at 1032 and 1038~\AA\ from the hot WD in the atmosphere of the red giant \citep{Schmid1989}. We did not detect these features in our spectra, consistent with the classification of \GR\ as an accreting-only system.

}

The rapid photometric flickering seen in the \TESS\/ light curve of \GR\ (Figure~\ref{fig:GRCyg_TESS}) is a signature of ongoing accretion in the system. At first it seems surprising that accretion-related flickering would be conspicuous at the relatively long wavelengths of the \TESS\/ bandpass ($\sim$6000--10000~\AA), where flux from the red giant dominates over that due to the accretion disk. However, in a recent study of \TESS\/ photometry of symbiotics, \citet{Merc2024} detected flickering in 80\% of a sample of 27 accreting-only systems.

%\clearpage

\section{Synthetic Narrow-band Imaging of \GR \label{sec:GRCyg_synthetic_imaging} }

Since LRS2-B is an IFU spectrograph, its resulting data cube can be sliced to create  spatial images, at any chosen wavelength. We used this capability to construct synthetic narrow-band emission-line-minus-continuum difference images of \GR\ from our LRS2-B observations. See \citet{BondAbell572024} for details of our procedures for creating this emission-line imaging. 

The small FOV of the LRS2-B IFU created difficulties in obtaining optimal data for \GR. We were interested especially in its strong [\oii] $\lambda$3727 emission, since our initial observations in 2025 suggested that this emission is displaced from the stellar spectrum. Due to atmospheric refraction, as well as this spatial displacement, the [\oii] emission fell near the edge of the FOV in our initial attempts in 2026 April. After several adjustments of the observation specifications, we eventually obtained a successful set of LRS2-B exposures (see Table~\ref{tab:GRCygLRSexposures} for a history of these attempts).

We combined the three exposures obtained on 2026 May~13 into a single stacked data cube. These observations had been acquired with a deliberate $2\farcs25$ eastward pointing offset relative to the continuum source, in order to place the previously identified displaced [\oii] emission near the center of the LRS2 FOV, while retaining the continuum source within the IFU\null. Individual exposures were reduced using the standard {\tt Panacea} pipeline (see Section~\ref{sec:HET_obs_of_GRCyg}), registered to a common astrometric frame using the continuum emission, and combined with inverse-variance weighting. Differential atmospheric refraction was corrected before constructing narrow-band images at individual wavelengths, so that all wavelengths are referenced to the astrometric frame defined by the 3780~\AA\ continuum centroid.

To isolate the faint emission lines from the much brighter stellar spectrum, we first modeled and removed the stellar continuum using carbon-star spectra (which lack emission lines) from the MaStar Stellar Library \citep{Yan2019}. The model was allowed to vary smoothly in overall shape to account for differences between the library spectra and our observations. We then constructed an image of each emission line by combining the residual signal over the wavelength range appropriate for that line. The weighting was matched to the spectral resolution of LRS2-B, giving the greatest weight to wavelengths nearest the line center and improving the sensitivity to faint, unresolved emission

\begin{figure*}
\centering
\includegraphics[width=5in]{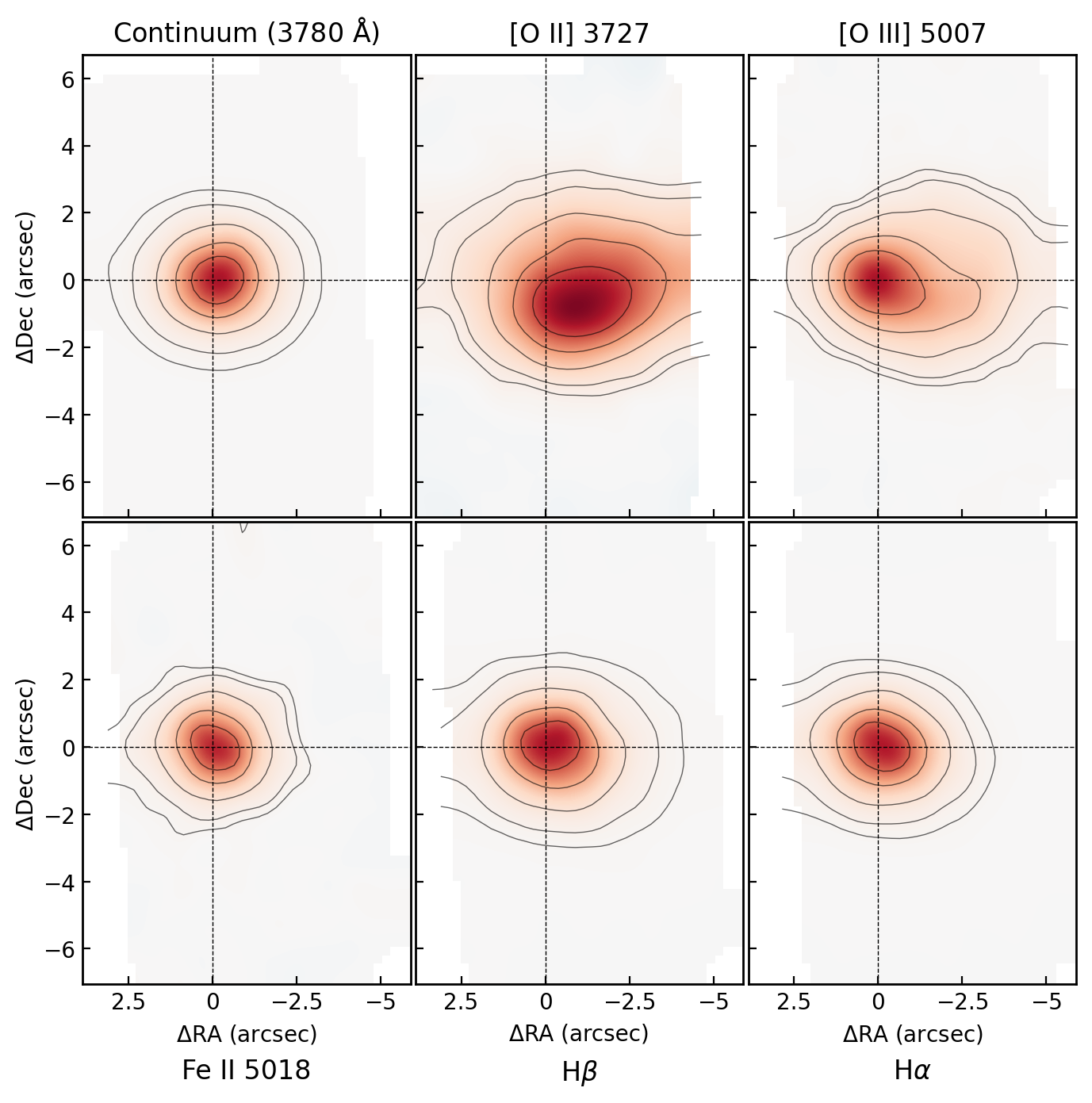}
%\vskip-0.4in
\caption{
Continuum (upper-left panel) and narrow-band synthetic emission-line images of GR~Cyg in [\oii], [\oiii], \feii, \Hb, and \Ha, constructed from the stacked HET/LRS2-B exposures obtained on 2026 May~13. Full width at half maximum of the continuum image is $1\farcs8$.
Origin of coordinates is the centroid of the 3780~\AA\ continuum image, corrected for differential atmospheric refraction. Contours show the regions containing 25\%, 50\%, 75\%, 90\%, and 95\% of the total light.The images in \feii\ and Balmer lines are stellar and coincident with the stellar continuum. However, the image in [\oii] $\lambda$3727 is elongated and offset from the stellar continuum by about $1\farcs6$, and [\oiii] $\lambda$5007 is elongated in about the same direction. 
\label{fig:GR Cyg_narrowband}
}
\end{figure*}

Figure~\ref{fig:GR Cyg_narrowband} presents the 3780~\AA\ continuum image of \GR\ in the upper-left panel. The rest of the panels show narrow-band images in the emission lines of [\oii] $\lambda$3727, [\oiii] $\lambda$5007, \feii\ $\lambda$5018, \Hb, and \Ha.  Spatial coordinates are shown relative to the location of the continuum centroid.

The \feii\ and Balmer-line images appear stellar at the resolution of these frames, and are spatially coincident with the continuum. However, the [\oii] $\lambda$3727 emission is partially resolved into an elongated structure. Its centroid is displaced by approximately $1\farcs6$ to the southwest of the continuum source, corresponding to $\sim$2900~AU at the distance of \GR. [\oiii] $\lambda$5007 shows a weaker elongation in the same direction.

Collimated jets and outflows have been detected around several symbiotic binaries, and the [\oii] feature in \GR\ is likely a new example. The symbiotic system CH~Cygni shows [\oii]-emitting nebulosities within several thousand AU of the star. These are fairly similar to the feature in \GR; see, for example, Figure~1 in \citet{Corradi2001}, which shows both ground-based and high-resolution {\it Hubble Space Telescope\/} (\HST) narrow-band images of CH~Cyg.

These outflows are often 
expanding along trajectories and
at velocities implying violent ejection processes
\citep[e.g.,][]{2001ApJ...553..211C, 
2007A&A...465..481S, 2025A&A...704L..11L}.  Our current observations of \GR, however, are
still incomplete for full characterization of the dynamical status of its outflow.

%\clearpage 

\null\bigbreak
\bigbreak

\section{Spectroscopy of \D141}

\subsection{Hobby-Eberly LRS2-B Observations \label{sec:HET_obs_of_D141} }

As in the case of \GR\ (see Section~\ref{sec:GRCyg_spectroscopy}), the unusual situation of a faint ionized nebula surrounding a late-type variable star prompted us to add \D141\ to the queue for our HET/LRS2-B spectroscopic survey program. Our initial observation was obtained in 2025 July, and since then two more LRS2-B spectra have been obtained.  An observing log for our LRS2-B spectroscopy is presented in Table~\ref{tab:D141_Obs_log}. The table also includes two observations obtained at Asiago, discussed below.

\begin{deluxetable}{lcc}[h]
\tablecaption{Log of HET and Asiago Spectroscopic Observations of \D141
\label{tab:D141_Obs_log} }
\tablehead{
\colhead{Telescope}
&\colhead{Date}
&\colhead{Exposure}\\
\colhead{}
&\colhead{[YYYY-MM-DD]}
&\colhead{[s]}}
\decimals
\startdata
HET & 2025-07-20 & $38   $	    \\
HET & 2025-08-29 & $38   $	    \\
Asiago 1.22 m & 2026-01-20 & 2200 \\
Asiago 1.82 m & 2027-02-01 & 900 \\
HET & 2026-05-08 & $98 $ \\      
\enddata
\end{deluxetable}

\begin{figure*}[!]
\centering
\includegraphics[width=6in]{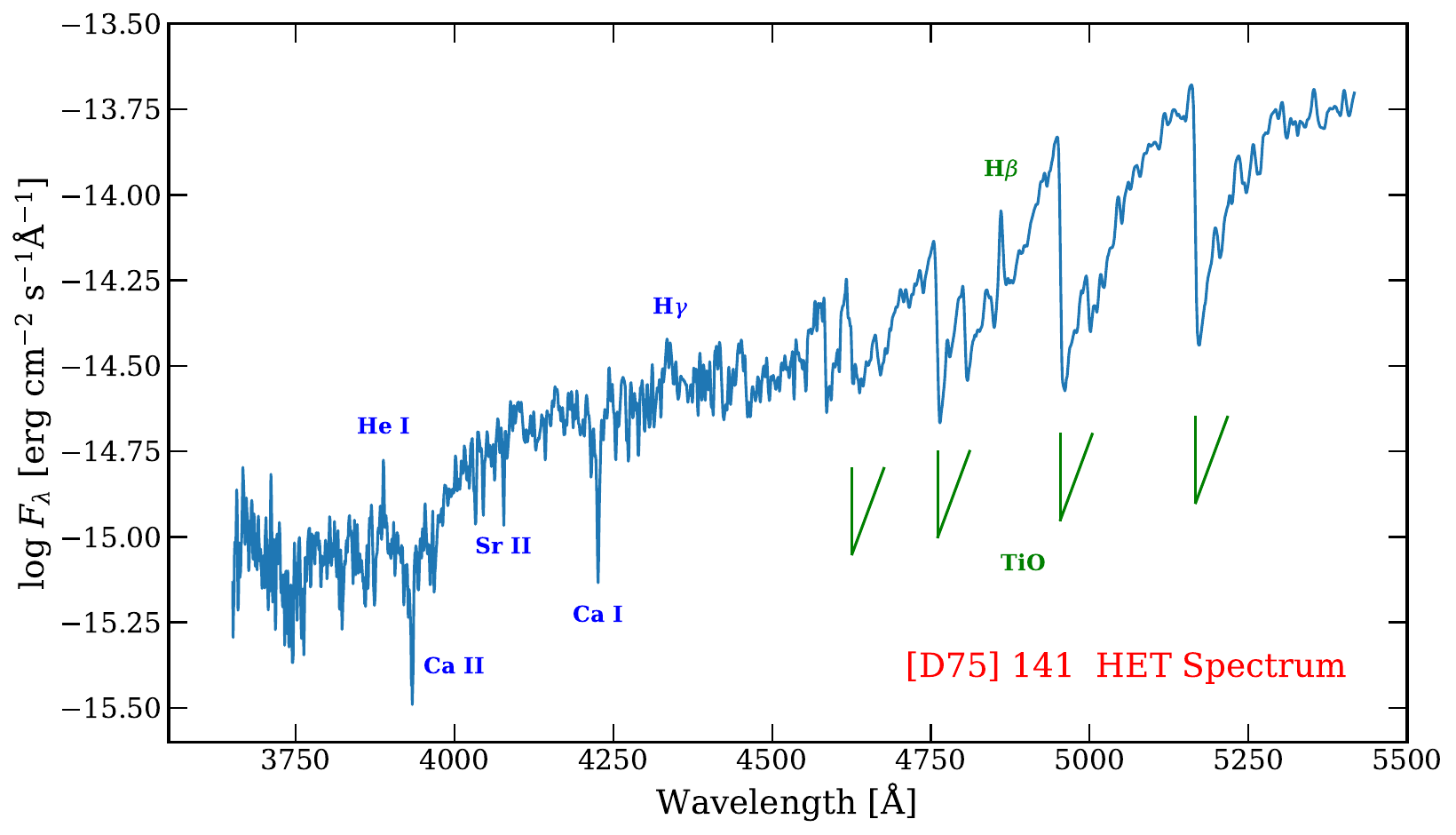}
%\vskip-0.4in
\caption{
Spectrum of \D141\ obtained with the HET LRS2-B spectrograph. Emission lines of the Balmer series are marked, along with TiO absorption bands and absorption lines of \cai, \caii, and \srii\ $\lambda$4077. \label{fig:D141_HETspectrum}
}
\end{figure*}

Figure~\ref{fig:D141_HETspectrum} plots a portion of the LRS2-B spectrum of \D141, from 3640 to 5400~\AA\null. The spectrum shown is a combination of all three exposures from 2025 and 2026. There were no significant variations seen among the three spectra, apart from minor changes in the intensity of Balmer emission lines. The spectrum shows strong absorption bands of TiO, consistent with the M spectral type reported by  \citet{Dolidze1975}. Weak emission is detected at \Hb\ and \Hg, as typically seen in late-type pulsating variable stars. Emission at \hei\ $\lambda$3889 is also present; the spectrum is noisy at these blue wavelengths, but the detection appears to be real. There are no forbidden emission lines seen in the spectrum, unlike the situation in \GR. However, as in the case of \GR, there is an increase in the flux level at the blue end of the spectrum, indicative of a hot companion and/or accretion disk. The absorption line of \srii\ $\lambda$4077, labeled in Figure~\ref{fig:D141_HETspectrum}, is notably strong and might suggest an excess of \sprocess\ elements, but a high-resolution abundance analysis would be needed to confirm this.

We prepared synthetic narrow-band images of \D141\ at \Ha, \Hb, and \hei\ $\lambda$3889,  as was done for \GR\ (Section~\ref{sec:GRCyg_synthetic_imaging}). These images proved to be stellar (at the $\sim$$1\farcs5$ seeing of the data), and showed the emission-line region to be spatially coincident with the continuum source.

\subsection{Asiago Observations}

As was the case for \GR, co-author Munari obtained spectra of \D141\ at Asiago  in 2026 January and February, just before the star approached solar conjunction. Details of the spectroscopic observations are included in Table~\ref{tab:D141_Obs_log}. The observation on 2026 February~1 was made on the same night as \GR\ was observed, with the
Asiago 1.82~m telescope and its REOSC-Echelle spectrograph (resolving power $R\simeq22,000$,
wavelength range 3550--7110~\AA).
As for \GR, the Echelle spectrum will be further investigated
in a forthcoming paper giving further details, and we focus here on the lower-resolution spectrum obtained with the 1.22~m telescope

The Asiago 1.22~m spectrum of [D75]~141 was
recorded using the same
telescope and spectrograph setup as for \GR, but with a different grating: a 300~line\,mm$^{-1}$
grating blazed at 5000~\AA, which allowed us to cover the 3300--8000~\AA\
range at 2.31~\AA\,pix$^{-1}$.  The long slit was again used, oriented
east-west with a width of $1\farcs8$.

Reductions and calibrations were done as described for \GR. A photometric observation was made on the same night using the Asiago 67/92 cm Schmidt, giving the results presented above in Table~\ref{tab:photometry}, and it was used to refine the absolute flux calibration of the 1.22~m spectrum.

\bigbreak

\subsection{Spectral Classification of \D141}

\begin{figure*}
\centering
\includegraphics[width=6in]{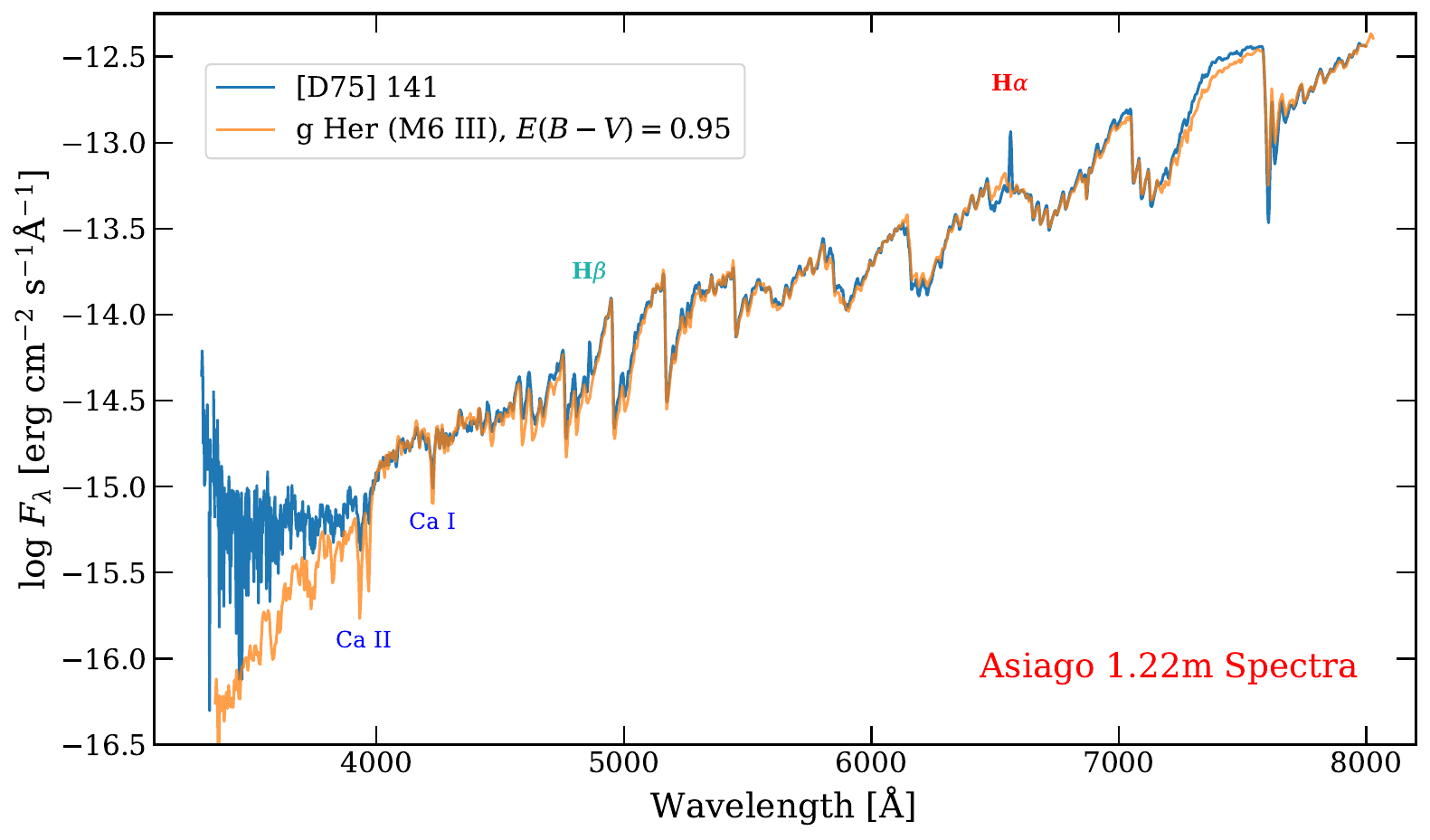}
%\vskip-0.4in
\caption{
Asiago 1.22~m spectrum of \D141\ (blue line) compared with that of the M6~III standard star g~Her (orange line). A reddening of $E(B-V)=0.95$ has been applied to g~Her. Emission lines of the \Ha\ and \Hb, and absorption lines of \cai\ and \caii\ are marked. Note the blue excess of \D141\ at the short-wavelength end. See text for discussion. 
\label{fig:D141_Asiagospectrum}
}
\end{figure*}

Figure~\ref{fig:D141_Asiagospectrum} plots (blue line) the Asiago 1.22~m spectrum of \D141. We compared this spectrum with a range of M-type standard stars obtained with the same instrumentation. We found a nearly perfect match (orange line) with that of the bright star g~Her (HD\,148783), an MKK standard for the M6~III spectral type \citep{Keenan1989}. To achieve a match to the SED of \D141, we had to apply a reddening of
$E(B-V) = 0.95$ to the standard star. As in the case of \GR\ (Section~\ref{sec:GRCyg_spectral_classification}), this reddening follows the $R_V = 3.1$ law of \citet{1999PASP..111...63F} and is in addition to the small amount
affecting g~Her. 

%\vskip0.2in\null
%\bigbreak

\subsection{\D141: Another Newly Discovered Symbiotic Binary}

The presence of a hot companion to the M6\,III
red giant in \D141\ is revealed by the near-UV excess at $\lambda \le 4000$~\AA\ shown in Figures~\ref{fig:D141_HETspectrum} and~\ref{fig:D141_Asiagospectrum}, and by
the Balmer and \hei\ emission lines.   This configuration is remarkably similar to that
of SU~Lyncis, a symbiotic system which likewise combines an M6\,III red giant with a hot companion, discovered by \citet{Mukai2016}. The close similarity is seen by comparing our Figure~\ref{fig:D141_Asiagospectrum} spectrum of \D141\ to that of SU~Lyn shown in Figure~4 of \citet{Mukai2016}. 
SU~Lyn is a prototype of the low-luminosity, accreting-only symbiotic stars, discussed in Section~\ref{sec:GRCyg_new_symbiotic}. 
We conclude thaat [D75]~141 is a newly discovered symbiotic star
of the same accreting-only class as SU~Lyn---as is also \GR.

\section{Discussion}

%Two new symbiotic binaries. There are several recent papers on the ``missing'' symbiotics (Mukai?). Can we speculate on how these two were missed for so long? Quick literature search indicates lots of new symbiotics have been discovered in past few years (but not these two!)

\subsection{\GR\ / \StDr \label{sec:GRCyg_discussion} }

GR~Cyg is a unique
symbiotic star, in the sense that it is both a carbon red giant, and it lies within an extended
ionized nebula. As discussed in Section~\ref{sec:GRCyg_new_symbiotic}, carbon stars are very rare among Galactic symbiotic binaries. Moreover, less than half of known dusty  symbiotics like \GR\ are surrounded by optical nebulae \citep[e.g.,][]{Ilkiewicz2018}.

{

As we noted in Section~\ref{sec:GRCyg_variability}, and as shown in Figure~\ref{fig:GRCyg_lightcurve}, the light curve of \GR\ shows irregular variations on top of the pulsations. The light curves of carbon-rich Miras are indeed known to be more erratic than those of the oxygen-rich ones (e.g., \citealt{Whitelock1997}). This has been attributed to ejection of C-rich dust, a phenomenon seen in the R~CrB variables. Additionally, the light curves of Miras in symbiotic binaries are less regular than those of single Miras (see the review by \citealt{Whitelock2003}).

}

A double-ring ``bull's-eye'' low-excitation ionized nebula surrounding a late-type symbiotic star is, to our knowledge, unprecedented. However, the symbiotic binary AS~201 is encircled by a single ionized ring, discovered by \citet{Schwarz1991}. This feature shares the low excitation level (strong [\nii], very weak [\oiii]) seen in \StDr, and the linear size of the ring is similar to that of the inner ring around \GR. The luminous M-type supergiant W26 in the massive young cluster Westerlund~1 is surrounded by an ionized ring with a radius of $\sim$0.05~pc \citep{Wright2014}. The origin of this ring's ionization remains unclear; it could be due to UV radiation from nearby hot stars in the cluster, an unseen hot companion of W26 itself, or collisional excitation in a wind from the star.

Among ionized true PNe surrounding hot stars, examination of high-resolution images shows that some $\sim$8\% of them are surrounded by rings or partial rings and arcs \citep{Ramos-Larios2016}. The number of rings ranges from one or a few, up to 10 or more; see, for example, the multiple nested rings in images of the PNe IC~418 and NGC~6543 presented by \citet{Montoro-Molina2025}. However, the multiple rings in these PNe are generally morphologically distinct from those around \GR, since they are a tightly spaced series of rings, rather than the widely spaced pair seen in \GR. Thus it is unclear whether these rings in PNe are evolutionarily related to those surrounding \GR.

%the PN HDW~3 bears some resemblance to \StDr, showing partial segments of two concentric arcs in the direction of its proper motion,\footnote{See the images of HDW~3 at \url{https://app.astrobin.com/i/ke08wr} and \url{https://www.imagingdeepspace.com/hdw3-planetary-nebula-perseus.html}} but they do not nearly completely encircle the central star like \StDr\ does. 

Detached cold molecular rings are seen in submillimeter imaging of many AGB stars, but only when they are carbon-rich; see, for example, the CO radio maps of  carbon stars in Figures~9 and~10 in the review article by \citet{Hofner2018}, and Figure~1 in the review by \citet{Decin2020}. These structures are sometimes smooth and nearly perfect circles similar to those in the ionized gas surrounding \GR. A set of smooth elliptical rings is detected in CO maps of the carbon star V~Hydrae \citep{Sahai2022}. We speculate that the rings around \GR\ may have a similar origin, but the presence of a hot companion has led to them being ionized instead of molecular.

The bull's-eye structure around \GR\ is suggestive of periodically recurring ejections. Two possible recurrent events are the helium-shell thermal pulses that occur in late AGB evolution and lead to episodes of enhanced mass-loss rates; and  nova outbursts triggered when hydrogen-burning eventually ignites on the surface of the accreting WD\null. The angular radii of the two rings in \StDr\ differ by about $19\farcs1$ or 0.17~pc (see Section~\ref{sec:GRCygDeepImaging}). If we suppose that the rings are expanding at a velocity of $v_{\rm exp}\simeq15\,\kms$, a typical escape velocity from an AGB star, the time interval between the two events is given by
\[ \Delta t \simeq 10900 \left({\Delta r\over{19\farcs1}}\right) \left({{15\,\kms}\over{v_{\rm exp}}}\right) \,\rm yr \, , \]
for an assumed distance of $\sim$1800~pc (Table~\ref{tab:GRCyg_DR3data}). This timescale is in reasonable agreement, within the uncertainties, with the time intervals between thermal pulses in AGB stars derived in theoretical modeling \citep[e.g.,][and references therein]{Karakas2014}. It was argued more than three decades ago \citep{Olofsson1990} that the detached cold circumstellar envelopes seen in CO radio maps of carbon stars are a natural consequence of mass loss during helium shell flashes. Optical imaging with \HST\/ has shown spherically symmetric thin dust shells around several carbon stars \citep[e.g.,][]{Olofsson2010}

Alternatively, it remains plausible that the rings around \GR\ are the result of recurring nova outbursts. In this case, the expansion velocity could be considerably higher than assumed in the above equation, making the recurrence interval correspondingly shorter. 
This nuclear fusion can even be explosive if
the WD is massive (as in the recurrent nova RS~Oph), or steady and slow in thermal
equilibrium over decades or centuries if the WD has a low mass (as
in V4368~Sgr or V1016~Cyg).  Upon exhaustion of the shell fuel, the nuclear
burning ceases and accretion resumes, preparing the system for a new cycle.

{

We investigated the historical record of \GR\ in the data base of the Digital Access to a Sky Century @ Harvard (DASCH; \citealt{Grindlay2012}) project.\footnote{\url{https://dasch.cfa.harvard.edu/dr7}} We found 110 detections of the star, based on plates obtained between 1890 and 1990. The data show that (1)~the present-day variability depicted in Figure~\ref{fig:GRCyg_lightcurve} was present at similar amplitudes and mean brightness throughout the historical record; and (2)~no bright outburst was detected on the Harvard plates. However, the gaps in coverage are such that a short nova eruption could have gone unobserved.

}

\subsection{\D141\ / StDr 4}

The nebula StDr~4 surrounding the symbiotic binary \D141\ clearly contains a bow shock, as shown by a morphology consistent with the direction of motion of the star, and with the strong [\sii] emission indicating collisional excitation. The bow shock is embedded within a larger ionized nebula.

We speculate that StDr~4 is the result of the passage of a symbiotic binary through a cloud in the ISM\null. Figure~\ref{fig:StDr4_colorimage} shows that the low-Galactic-latitude field surrounding StDr~4 is covered with faint ambient nebulosity. Our picture is that an outflow from \D141, generated either by a thermal pulse or a symbiotic nova outburst, is colliding with the ISM\null. Further outside the shock, UV radiation from the accreting component of the binary photoionizes the interstellar gas---or, alternatively, this larger nebulosity could be the partially dissipated remnant of a previous ejection event.

\section{Summary and Future Work}

We have been conducting a spectroscopic survey of stars surrounded by
faint nebulae discovered by amateur astronomers. A large majority of our targets prove to be PNe, with hot post-AGB nuclei. In this paper, however, we investigate two cases in which the central stars are unusually cool, and the nebulae are not classical PNe. We present extremely deep narrow-band images of both nebulae, revealing remarkable morphologies. The \StDr\ nebula exhibits a pair of concentric ionized rings, while StDr~4 shows a bow shock embedded in a larger diffuse nebula. Photometric monitoring surveys show that both central stars---\GR\ and \D141---are late-type semiregular variables. \GR\ is a carbon star, while \D141\ is an M6 red giant. We obtained spectroscopy demonstrating that the two stars are previously unrecognized symbiotic binaries, combining a late-type red giant with an accreting hot companion. Both binaries appear to belong to the subclass of accreting-only symbiotics. 

The fact that these two relatively bright symbiotics were revealed only because they happened to be surrounded by faint nebulosities discovered by amateurs supports the view that a significant population of ``missing'' symbiotics remains to be discovered (see, for example, \citealt{Mukai2016}, \citealt{Munari2021}, \citealt{Xu2024}, \citealt{Merc2025}, and \citealt{Contreras2026}). \GR\ is a very rare example of a carbon-rich symbiotic binary in the Milky Way. In addition to the concentric rings around \GR, there is a compact [\oii]-emitting knot lying only $1\farcs6$ away from the star; this feature likely has a different physical origin from the larger rings.

Future studies of these objects would be useful. These include (1)~high-resolution imagery, from the ground or preferably from space, including measurements of the angular expansion of the nebulae; (2)~UV spectroscopy to characterize the hot companions of the red-giant components; (3)~high-resolution optical spectroscopy, both for abundance studies and to determine orbital periods of the binaries.

%\clearpage 

\acknowledgments

We thank the referee for useful comments which improved our presentation.

We thank the HET queue schedulers and nighttime observers at McDonald Observatory for obtaining the LRS2-B data discussed here---and especially for the tricky placement of \GR\ in the FOV.

The Low-Resolution Spectrograph 2 (LRS2) was developed and funded by The University of Texas at Austin McDonald Observatory and Department of Astronomy, and by The Pennsylvania State University. We thank the Leibniz-Institut f\"ur Astrophysik Potsdam (AIP) and the Institut f\"ur Astrophysik G\"ottingen (IAG) for their contributions to the construction of the integral-field units.

We acknowledge the Texas Advanced Computing Center (TACC) at The University of Texas at Austin for providing high-performance computing, visualization, and storage resources that have contributed to the results reported within this paper.

This work has made use of data from the European Space Agency (ESA) mission
{\it Gaia\/} (\url{https://www.cosmos.esa.int/gaia}), processed by the {\it Gaia\/} Data Processing and Analysis Consortium (DPAC,
\url{https://www.cosmos.esa.int/web/gaia/dpac/consortium}). Funding for the DPAC
has been provided by national institutions, in particular the institutions
participating in the {\it Gaia\/} Multilateral Agreement.

Based in part on observations made with the NASA {\it Galaxy Evolution Explorer}.
\GALEX\/ was operated for NASA by the California Institute of Technology under NASA
contract NAS5-98034. \GALEX\/ data were obtained from the MAST data archive at the Space Telescope Science Institute, which is operated by the Association of Universities for Research in Astronomy, Inc., under NASA contract NAS5-26555.

Funding for the \TESS\/ mission is provided by NASA's Science Mission directorate.

This research has made use of the SIMBAD database, operated at CDS, Strasbourg, France \citep{Wenger2000}.

This research has made use of the VizieR catalogue access tool, CDS,
Strasbourg, France \citep{10.26093/cds/vizier}. The original description 
of the VizieR service was published in \citet{vizier2000}.

The Digitized Sky Surveys were produced at the Space Telescope Science Institute under U.S. Government grant NAG W-2166. The images of these surveys are based on photographic data obtained using the Oschin Schmidt Telescope on Palomar Mountain and the UK Schmidt Telescope. The plates were processed into the present compressed digital form with the permission of these institutions. 

This work has made use of data provided by Digital Access to a Sky Century @ Harvard (DASCH), which has been partially supported by NSF grants AST-0407380, AST-0909073, and AST-1313370. Work on DASCH Data Release 7 received support from the Smithsonian American Women's History Initiative Pool. We thank P.~Valisa and A.~Milani (ANS Collaboration) for their assistance in accessing the DASCH database.

\bibliography{PNNisurvey_refs}

\end{document}